\documentclass[aps,floatfix,twocolumn,nofootinbib,superscriptaddress,preprintnumbers]{revtex4-2} 

\usepackage{graphicx}
\usepackage{comment}
\usepackage{dcolumn}
\usepackage{bm}
\usepackage{physics}
\usepackage{amsmath,amssymb,color,float,mathrsfs} 
\usepackage{xparse,etoolbox}
\usepackage{etaremune}
\usepackage[capitalize]{cleveref}

\newcommand{\iu}{{i\mkern1mu}} 
\newcommand{\hatn}{\hat{\bm{n}}}
\newcommand{\rmd}[2]{\,{\rm d}^{#1} #2 \,} 
\newcommand{\Int}[3]{%
	\ifstrempty{#3}%
	{\int \!\! \rmd{#1}{#2} \,}%
	{\int \!\! \frac{\rmd{#1}{#2}}{#3} \,}%
}%
\newcommand{\INT}[5]{%
	\ifstrempty{#3}%
	{\int_{#4}^{#5} \!\! \rmd{#1}{#2} \,}%
	{\int_{#4}^{#5} \!\! \frac{\rmd{#1}{#2}}{#3} \,}%
}%

\def\apj{Astrophys.\ J.\ }
\def\apjl{Astrophys.\ J.\ Lett.\ }
\def\jcap{J.\ Cosmol.\ Astropart.\ Phys.\ }

\def\mnras{Mon.\ Not.\ R.\ Astron.\ Soc.\ }

\def\prd{Phys.\ Rev.\ D }
\def\plb{Phys.\ Lett.\ B }
\def\prl{Phys.\ Rev.\ Lett.\ }

\def\ptep{Prog.\ Theor.\ Exp.\ Phys.\ }

\begin{document}

\preprint{RESCEU-12/23}

\title{Gravitational lensing effect on cosmic birefringence}

\author{Fumihiro Naokawa}
\affiliation{Department of Physics, Graduate School of Science, The University of Tokyo, Bunkyo-ku, Tokyo 113-0033, Japan}
\affiliation{Research Center for the Early Universe, The University of Tokyo, Bunkyo-ku, Tokyo 113-0033, Japan}
\author{Toshiya Namikawa}%
\affiliation{Center for Data-Driven Discovery, Kavli IPMU (WPI), UTIAS, The University of Tokyo, Kashiwa, 277-8583, Japan}%


\begin{abstract}
We calculate the effect of gravitational lensing on the parity-odd power spectrum of the cosmic microwave background (CMB) polarization induced by axionlike particles (ALPs). Several recent works have reported a tantalizing hint of cosmic birefringence, a rotation of the linear polarization plane of CMB, which ALPs can explain. In future CMB observations, we can measure cosmic birefringence more precisely to get insight into ALPs. We find that the lensing effect is necessary to fit the observed $EB$ power spectrum induced by cosmic birefringence in future CMB observations, including Simons Observatory and CMB-S4. We also show that the estimated ALPs parameters are biased if we ignore the lensing effect. Therefore, the lensing correction to the parity-odd power spectra must be included in future high-resolution CMB experiments. 
\end{abstract}

\maketitle


\section{\label{sec:Intro}Introduction}
{\it Cosmic birefringence} -- a rotation of the linear polarization plane of cosmic microwave background (CMB) as they travel through space -- is of great interest in recent cosmology \cite{Komatsu:2022:review}. 
Cosmic birefringence can be caused by axionlike particles (ALPs), $\phi$, pseudoscalar fields that interact with photons via the so-called Chern-Simons term: ${\cal L}\supset -g\phi F^{\mu\nu}\tilde{F}_{\mu\nu}/4$ \cite{Carroll:1998zi}. Here, $g$ is the photon-axion coupling constant, $F^{\mu\nu}$ is the electromagnetic tensor, and $\tilde{F}^{\mu\nu}$ is the Hodge dual of $F^{\mu\nu}$. During the propagation of photons through space filled with ALPs, $\phi$ changes depending on cosmology and their potential. The net rotation angle is then given by \cite{Carroll:1989vb,Carroll:1991zs,Harari:1992ea}
\begin{equation}
    \label{23051201}
    \beta = \frac{1}{2}g \int_{t_{\rm LSS}}^{t_0}\frac{{\rm d}\phi}{{\rm d}t}{\rm d}t 
    \,,
\end{equation}
where $t_{\rm LSS}$ and $t_0$ are the time at the last scattering and the present day. This rotation leads to non-zero parity-odd power spectra, the temperature-$B$, and $EB$ correlations \cite{Lue:1998mq,Feng:2004mq}. By measuring $\beta$ from these parity-odd spectra, we can constrain ALPs models, which determines the evolution of $\phi$ \cite{Fujita:2020ecn}. ALPs can act as dark matter \cite{Finelli:2008,Fedderke:2019ajk} or dark energy \cite{Carroll:1998zi,Panda:2010uq} and is a possible signature of quantum gravity \cite{Myers:2003fd,Arvanitaki:2009fg}. Therefore, constraining the physical parameters of ALPs can greatly impact physics.

Several studies have recently reported a tantalizing hint of cosmic birefringence using the latest CMB polarization data \cite{Minami:2020odp,Diego-Palazuelos:2022dsq,Eskilt:2022cff}. 
The latest value of $\beta$ is $0.342^{+0.094}_{-0.091}~\mathrm{deg}$ ($1\,\sigma$), which excludes $\beta=0$ at $3.6\,\sigma$ level \cite{Eskilt:2022cff}. Constraining $\beta$ with higher statistical significance to confirm a nonzero $\beta$ will be one of the main goals in ongoing and future CMB experiments such as LiteBIRD \cite{LiteBIRD:2022:PTEP}, Simons Observatory (SO) \cite{SimonsObservatory}, and CMB-S4 (S4) \cite{CMBS4:r-forecast}. 
However, these observational results assume a constant rotation angle. This assumption is invalid if $\phi$ evolves during generating polarization signals. The evolution of $\phi$ can significantly modify the shape of the $EB$ power spectrum \cite{Finelli:2008,Sherwin:2021,Nakatsuka:2022epj,Galaverni:2023}. A precision measurement of $EB$ spectral shape, thus, provides detailed information on the dynamics of $\phi$, especially during recombination and reionization epoch. 

This paper addresses a crucial effect on the ALPs-induced cosmic birefringence --- the gravitational lensing effect on CMB. The lensing effect perturbs the trajectory of CMB photons, distorts the small-scale fluctuations of the CMB polarization map, and converts part of $E$-modes into $B$-modes (see \cite{Lewis:2006fu,Hanson:2009kr,Namikawa:2014xga} for review). Multiple CMB observations have detected the lensing effect on the CMB angular power spectra \cite{Keisler:2011aw,ACT:Choi:2020,SPT-3G:2021,Rosenberg:2022:PR4}, which will be detected with higher statistical significance in the future. In near-term and future CMB observations, we need to include the lensing effect in the theoretical calculation of the CMB power spectra to constrain cosmology. 

Several previous works, including Refs.~\cite{Sherwin:2021,Nakatsuka:2022epj}, have ignored the lensing effect on the calculation of the cosmic birefringence when $\phi$ evolves and the rotation angle has time dependence. 
Note that Ref.~\cite{Gubitosi:2014cua} included the lensing effect in their calculation of the CMB power spectra assuming that $\beta$ evolves linearly in the conformal time, although their model leads to a negligible time evolution during recombination. In this paper, we solve the time evolution of $\phi$ to accurately predict the ALPs-induced cosmic birefringence with the lensing effect. Calculating an accurate $EB$ power spectrum for a given ALPs model, we show the importance of the lensing effect on ALPs search with cosmic birefringence. 

This paper is organized as follows. Section \ref{Sec:biref} reviews the mechanism of cosmic birefringence and $EB$ power spectrum by ALPs. Section \ref{sec:lensing} reviews the lensing correction for the parity-even CMB power spectra following Ref.~\cite{Challinor:2005jy} and introduces the correction for the parity-odd power spectra. Section \ref{23042520} shows the $EB$ power spectrum with lensing correction and its significance in observations. Section \ref{23042521} shows the lensing effect on the ALPs search. Section \ref{sec:summary} is devoted to summary and discussion.

\section{Cosmic birefringence} \label{Sec:biref}
In this section, we follow Ref.~\cite{Nakatsuka:2022epj} to introduce cosmic birefringence and its impact on CMB observables, including the case where the approximation of the constant birefringence angle is not valid due to the time evolution of $\phi$. 

\subsection{constant rotation}
If the time evolution of $\phi$ is negligible during recombination, the linear polarization plane of all CMB polarization generated at the recombination is rotated by the same constant angle given in Eq.~\eqref{23051201}. The rotation angle is simply proportional to the difference between the field values at present and that at the recombination. The rotation of the linear polarization plane by a constant angle, $\beta$, then modifies the Stokes $Q$ and $U$ parameters observed at a line-of-sight direction, $\hatn$, as\footnote{Strictly speaking, the CMB polarization is also generated during reionization. If the field values at reionization are close to that at recombination, Eq.~\eqref{Eq:QU-rot} is still valid.}
\begin{equation}
    P^o(\hatn) = e^{2\iu\beta}P(\hatn) 
    \,, \label{Eq:QU-rot}
\end{equation}
where we define $P \equiv Q+\iu U$ and the subscript $o$ denotes observables after the rotation. The Stokes parameters are transformed to $E$- and $B$-modes as \cite{Zaldarriaga:1996xe,Kamionkowski:1996ks}
\begin{equation}
    E_{lm}\pm\iu B_{lm} = - \Int{2}{\hatn}{} _{\pm 2}Y^*_{lm}(\hatn) [Q\pm\iu U] (\hatn)
    \,, \label{Eq:EB-def}
\end{equation}
where $_{\pm 2}Y_{lm}$ is the spin-$2$ spherical harmonics. Substituting Eq.~\eqref{Eq:QU-rot} into Eq.~\eqref{Eq:EB-def}, we obtain
\begin{align}
    E^o_{lm} &= E_{lm}\cos(2\beta) - B_{lm}\sin(2\beta)
    \,, \\ 
    B^o_{lm} &= E_{lm}\sin(2\beta) + B_{lm}\cos(2\beta)
    \,.
\end{align}
Consequently, the angular power spectra of CMB polarization are given by
\begin{align}
    C_l^{EE,o} &= C_l^{EE}\cos^2(2\beta) + C_l^{BB}\sin^2(2\beta)
    \,, \\
    C_l^{BB,o} &= C_l^{EE}\sin^2(2\beta) + C_l^{BB}\cos^2(2\beta)
    \,, \\
    C_l^{EB,o} &= \frac{1}{2}(C_l^{EE}-C_l^{BB})\sin(4\beta)
    \,. 
\end{align}
Here, we ignore the intrinsic correlation between $E$- and $B$-modes.
From these three relations, we obtain \cite{Zhao:2015mqa}
\begin{equation}
    \label{23041102}
    C_l^{EB,o}=\frac{1}{2}(C_l^{EE,o}-C_l^{BB,o})\tan(4\beta)
    \,.
\end{equation}
We can estimate $\beta$ from observations with the above equation. We note that, in reality, the overall instrumental miscalibration angle, $\alpha$, further rotates the polarization plane \cite{QUaD:2008ado,Miller:2009pt,Komatsu:2010:WMAP7,Keating:2013}. Thus, $\beta$ is degenerate with $\alpha$, and we need other polarization sources to calibrate $\alpha$, such as Galactic foregrounds \cite{Minami:2019ruj}, an astrophysical source \cite{Aumont:2018epb}, and an artificial polarization source \cite{Cornelison:2022zrc} to calibrate $\alpha$ and break this degeneracy.

\subsection{ALPs-induced cosmic birefringence}

Approximation of a constant angle is invalid if $\phi$ changes during the recombination or reionization epoch. For example, if values of $\phi$ at the recombination and reionization epochs differ, polarization generated at these epochs could be rotated by different $\beta$. To include the time evolution of $\phi$, we must simultaneously solve the Boltzmann equation for CMB polarization and the evolution equation of $\phi$. 

The Boltzmann equation for the CMB photons with cosmic birefringence is given by \cite{Liu:2006uh,Finelli:2008,Gubitosi:2014cua,Lee:2016jym} 
\begin{align}
    \label{23041101}
    &_{\pm2}\Delta'_P + \iu q\mu~_{\pm2}\Delta_P 
    \notag \\
    &\qquad = a n_{\rm e}\sigma_T
        \left[
            -~_{\pm2}\Delta_P + \sqrt{\frac{6\pi}{5}}~_{\pm2}Y_{20}(\mu)\Pi(\eta,q)
        \right]
    \notag \\
    &\qquad\qquad \pm \iu g\phi'~_{\pm2}\Delta_P
    \,. 
\end{align}
Here, $_{\pm2}\Delta_P$ is the Fourier coefficient of $Q\pm \iu U$ and is the function of the conformal time, $\eta$, the magnitude of the Fourier wavevector, $q$, and the cosine of the angle between the Fourier wavevector and line-of-sight direction, $\mu$. We also introduce the scale factor, $a$, the electron number density, $n_{\rm e}$, the cross-section of the Thomson scattering, $\sigma_T$, and the polarization source term, $\Pi$, introduced in Ref.~\cite{Zaldarriaga:1996xe}. The prime denotes the conformal time derivative. 
The evolution equation of $\phi$ is given by
\begin{equation}
    \phi'' + 2\frac{a'}{a}\phi' + a^2m_\phi^2 \phi = 0
    \,, \label{Eq:phi-EoM}
\end{equation}
for the ALP potential of $V(\phi)=m_\phi^2\phi^2/2$. 

To derive the CMB angular power spectra from $_{\pm2}\Delta_P$, we expand the Fourier coefficients as \cite{Zaldarriaga:1996xe}
\begin{align}
    _{\pm2}\Delta_P(\eta_0,q,\mu) 
    &\equiv -\sum_i \iu^{-l}\sqrt{4\pi(2l+1)} 
    \notag \\
    &\quad\times [\Delta_{E,l}(q)\pm \iu\Delta_{B,l}(q)] {}_{\pm2}Y_{l0}(\mu) 
    \,. 
\end{align}
The angular power spectra from these $E$- and $B$-modes are then given by
\begin{equation}
    C_l^{XY} = 4\pi
    \Int{}{(\ln q)}{} \mathcal{P}_s(q)\Delta_{X,l}(q)\Delta_{Y,l}(q)
    \,, \label{Eq:ClXY}
\end{equation}
where $\mathcal{P}_s(q)$ is the dimensionless power spectrum of the primordial scalar curvature, and $X$ and $Y$ denote either $E$ or $B$.
A solution of Eq.~(\ref{23041101}) provides the exact shape of $C_l^{EB}$ defined in Eq.~\eqref{Eq:ClXY}. 

Ref.~\cite{Nakatsuka:2022epj} developed a code, \texttt{birefCLASS}, by modifying the public Boltzman solver, \texttt{CLASS} \cite{CLASS}, to solve Eqs.~(\ref{23041101}) and (\ref{Eq:phi-EoM}). 
However, \texttt{birefCLASS} neglected the gravitational lensing effect, which substantially changes the shape of $C_l^{EB}$ at high-$l$ as we show in this paper.


\section{Gravitational Lensing Effect}
\label{sec:lensing}
We compute the gravitational lensing correction to the parity-odd power spectra. To derive the formula for the lensed parity-odd power spectra, we closely follow the curved-sky non-perturbative approach developed by Ref.~\cite{Challinor:2005jy}\footnote{See also the first non-perturbative calculation by Ref.~\cite{Seljak:1995} in the flay-sky approximation.} but keep the parity-odd power spectra in the calculation. 
We first review the derivation of the non-perturbative lensing correction to the temperature power spectrum. Then, we derive the lensing correction to the polarization power spectra, including parity-odd power spectra.

\subsection{temperature fluctuations}

\begin{figure}[t]
  \centering
  \includegraphics[width=8cm]{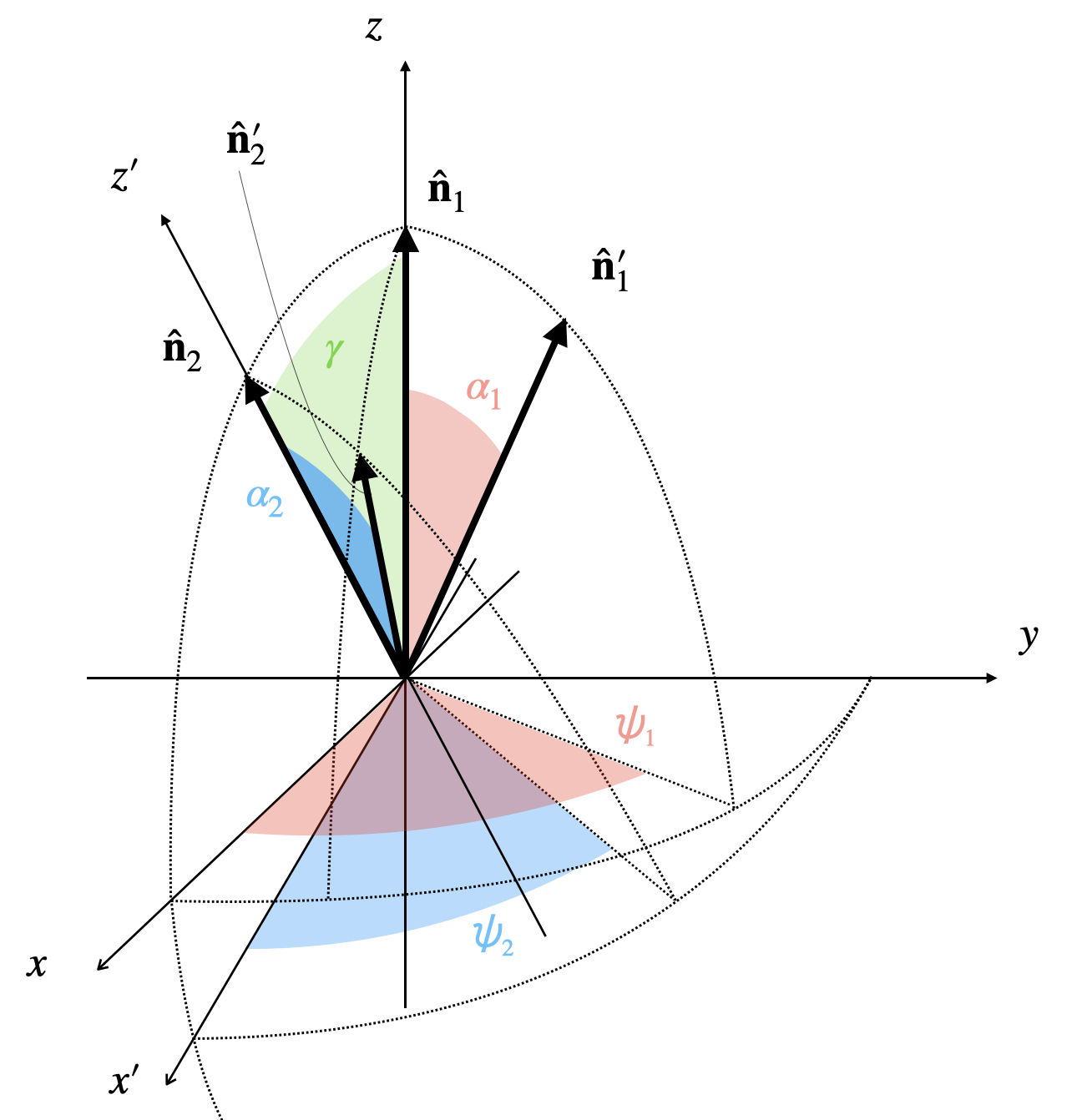}
  \caption{The coordinates we use for the lensing calculation of CMB angular power spectrum. The $x'$-$z'$ plane is obtained by rotating the $x$-$z$ plane around the $y$-axis by an angle $\gamma$. $\alpha_1$ and $\psi_1$ are the azimuthal and polar angles of $\hatn'_1$ in the $x$-$y$-$z$ coordinates, respectively, while $\alpha_2$ and $\psi_2$ are those of $\hatn'_2$ in the $x'$-$y$-$z'$ coordinates, respectively.}
  \label{fig:coordinate}
\end{figure}

The temperature power spectrum is generally written in terms of the two-point correlation function as 
\begin{equation}
    C_l^{\Theta\Theta} = 2\pi\INT{}{(\cos\gamma)}{}{-1}{1}
    \xi(\gamma)d_{00}^l(\gamma)
    \,. \label{Eq:xi}
\end{equation}
Here, $d_{mm'}^l(\gamma)$ is the Wigner $d$-function, $\xi(\gamma)$ is the correlation function for a given temperature map, $\Theta(\hatn)$, defined as $\xi(\gamma)=\ev{\Theta(\hatn_1)\Theta(\hatn_2)}$, and $\gamma$ is an angle\footnote{In this paper, we assign $\gamma$ instead of $\beta$ used in Ref.~\cite{Challinor:2005jy} to avoid confusion with the rotation angle $\beta$ of cosmic birefringence.} between two arbitrary unit vectors $\hatn_1$ and $\hatn_2$: $\cos\gamma\equiv\hatn_1\cdot\hatn_2$. The bracket, $\ev{\cdots}$, denotes the ensemble average. 

Let us include the lensing correction to Eq.~(\ref{Eq:xi}). We first define the coordinates as shown in Fig.~\ref{fig:coordinate}. We put $\hatn_1$ along the $z$-axis and $\hatn_2$ in the $x$-$z$ plane. When considering gravitational lensing, the CMB photons observed in the direction, $\hatn_1$, come from a displaced direction, $\hatn'_1=\hatn_1+\bm{\nabla}\Psi$, on the last scattering surface, where $\bm{\nabla}$ is the covariant derivative on the unit sphere, and $\Psi$ is the lensing potential. Similarly, the CMB photons observed in the direction, $\hatn_2$, come from $\hatn'_2=\hatn_2+\bm{\nabla}\Psi$ on the last scattering surface. The temperature fluctuations observed in $\hatn_i$ ($i=1,2$), $\tilde{\Theta}(\hatn_i)$, are equivalent to $\Theta(\hatn'_i)=\Theta(\hatn_i+\bm{\nabla}\Psi)$. The correlation function for the lensed temperature fluctuations is then given by
\begin{equation}
    \tilde{\xi}(\gamma)
    = \ev{\tilde{\Theta}(\hatn_1)\tilde{\Theta}(\hatn_2)}
    = \ev{\Theta(\hat{\bm{n}}'_1)\Theta(\hat{\bm{n}}'_2)}
    \,. 
\end{equation}
Here and after, we use a tilde for lensed quantities. Using the spherical harmonic transform, this lensed correlation function is expressed in terms of the unlensed power spectrum as
\begin{equation}
  \label{23041104}
    \tilde{\xi}(\gamma)
    =\sum_{lmm'}C_l^{\Theta\Theta} d_{mm'}^l(\gamma)
  \langle Y^*_{lm}(\alpha_1,\psi_1)Y_{lm'}(\alpha_2,\psi_2)\rangle
  \,, 
\end{equation}
where $Y_{lm}$ is the spherical harmonics, and $\alpha_1$, $\psi_1$, $\alpha_2$, $\psi_2$ characterize the directions of $\hatn'_1$ and $\hatn_2'$ as summarized in Fig.~\ref{fig:coordinate}. As shown in Ref.~\cite{Challinor:2005jy}, the last component, $A_l(\gamma)\equiv\langle Y^*_{lm}(\alpha_1,\psi_1)Y_{lm'}(\alpha_2,\psi_2)\rangle$, in Eq.~(\ref{23041104}) is expressed as a function of the lensing-potential power spectrum, $C_l^{\Psi\Psi}$ (see Eq.~(38) of Ref.~\cite{Challinor:2005jy} for details). 

Finally, we obtain the lensed power spectrum by substituting Eq.~(\ref{23041104}) into Eq.~(\ref{Eq:xi}):
\begin{equation}
    \label{23041103}
    \tilde{C}_l^{\Theta\Theta} 
    = 2\pi\sum_{l'mm'}\INT{}{(\cos\gamma)}{}{-1}{1}
    C_{l'}^{\Theta\Theta} d_{00}^l(\gamma)d_{mm'}^{l'}(\gamma)
    A_{l'}(\gamma)
\end{equation}
The lensed power spectrum is expressed as a weighted integration of the unlensed power spectrum.

\subsection{polarization}

Next, we derive the lensing correction to the parity-odd power spectra. 

We first describe the relationship between the correlation function and angular power spectra involving polarization, $C_l^{EE}$, $C_l^{BB}$, $C_l^{EB}$, $C_l^{\Theta E}$, and $C_l^{\Theta B}$. The correlation functions involving polarization are defined as \cite{Challinor:2005jy}:
\begin{align}
    \xi_+(\gamma) &\equiv \ev{P^*(\hatn_1)P(\hatn_2)}
    \,, \label{Eq:xi-pol1} \\
    \xi_-(\gamma) &\equiv \ev{P(\hatn_1)P(\hatn_2)}
    \,, \label{Eq:xi-pol2} \\
    \xi_X(\gamma) &\equiv \ev{\Theta(\hatn_1)P(\hatn_2)}
    \,. \label{Eq:xi-polx}
\end{align}
The relations between the above correlation functions and the power spectra are given by \cite{Chon:2003gx}:\footnote{Our convention of $U$ has the opposite sign to that in Ref.~\cite{Chon:2003gx}. This leads to the opposite sign in the imaginary part compared to Ref.~\cite{Chon:2003gx}.} 
\begin{align}
    &C_l^{EE} + C_l^{BB} = 2\pi\INT{}{(\cos\gamma)}{}{-1}{1} \xi_+(\gamma)d_{22}^l(\gamma)
    \,, \label{23041107} \\
    &C_l^{EE} - C_l^{BB} + 2\iu C_l^{EB} = 2\pi\INT{}{(\cos\gamma)}{}{-1}{1}
    \notag \\
    &\qquad\qquad\qquad\qquad\qquad\qquad\qquad \times \xi_-(\gamma)d_{2,-2}^l(\gamma)
    \,, \label{23041108} \\
    &C_l^{\Theta E} + \iu C_l^{\Theta B} = -2\pi\INT{}{(\cos\gamma)}{}{-1}{1}\xi_X(\gamma)d_{20}^l(\gamma)
    \,. \label{23041109}
\end{align}

Next, we compute the lensed correlation function. Substituting $\tilde{P}=\tilde{Q}\pm\tilde{U}$
into Eqs.~\eqref{Eq:xi-pol1}-\eqref{Eq:xi-polx} and employing Eq.~\eqref{Eq:EB-def} to transform the Stokes parameters to the $E$- and $B$-modes, we find
\begin{equation}
    \tilde{\xi}_+(\gamma)
        = \sum_{lmm'}(C_l^{EE}+C_l^{BB})d_{mm'}^l(\gamma)A_l^+(\gamma)
    \,, \label{23050201} 
\end{equation}
\begin{equation}
    \tilde{\xi}_-(\gamma)
        = \sum_{lmm'}(C_l^{EE}-C_l^{BB}+2\iu C_l^{EB}) d_{mm'}^l(\gamma)A_l^-(\gamma)
    \,, \label{23050202}    
\end{equation}
\begin{equation}
    \tilde{\xi}_X(\gamma) 
        = -\sum_{lmm'}(C_l^{\Theta E}+\iu C_l^{\Theta B})d_{mm'}^l(\gamma)A_l^X(\gamma)
    \,, \label{23050203} 
\end{equation}
where we keep parity-odd terms and define
\begin{align}
    A^+_l(\gamma) &\equiv \langle 
        e^{2\iu\psi_1 }_{~~~~+2}Y_{lm}(\alpha_1,\psi_1)_{~+2}Y^*_{lm'}(\alpha_2,\psi_2)e^{-2\iu\psi_2}
        \rangle
    \,, \\
    A^-_l(\gamma) &\equiv \langle 
        e^{-2\iu \psi_1 }~_{-2}Y_{lm}(\alpha_1,\psi_1)_{~+2}Y^*_{lm'}(\alpha_2,\psi_2)e^{-2\iu\psi_2}
        \rangle
    \,, \\
    A^X_l(\gamma) &\equiv \langle 
        Y_{lm}(\alpha_1,\psi_1)_{~+2}Y^*_{lm'}(\alpha_2,\psi_2)e^{-2\iu\psi_2}
        \rangle
    \,. 
\end{align}
The quantities, $A^+_l(\gamma)$, $A^-_l(\gamma)$, and $A^X_l(\gamma)$, are expressed in terms of the lensing-potential power spectrum whose explicit form is found in Eqs.~(54)-(56) of Ref.~\cite{Challinor:2005jy}. 

Finally, substituting the lensed correlation functions, $\tilde{\xi}_+$, $\tilde{\xi}_-$, and $\tilde{\xi}_X$, into Eqs.~(\ref{23041107})-(\ref{23041108}), we obtain the following expressions for the lensed power spectra: 
\begin{widetext}
\begin{align}
    \tilde{C}_l^{EE}+\tilde{C}_l^{BB} 
        &= 2\pi\sum_{l'mm'}\INT{}{(\cos\gamma)}{}{-1}{1}d^{l'}_{mm'}(\gamma)
        \left(C_{l'}^{EE}+C_{l'}^{BB}\right)A_{l'}^+(\gamma)d_{22}^l(\gamma)
    \,, \label{23041201} \\
    \tilde{C}_l^{EE}-\tilde{C}_l^{BB}+2\iu\tilde{C}_l^{EB}
        &= 2\pi\sum_{l'mm'}\INT{}{(\cos\gamma)}{}{-1}{1}d^{l'}_{mm'}(\gamma)
        \left( C_{l'}^{EE}-C_{l'}^{BB}+2\iu C_{l'}^{EB}\right)
        A_{l'}^-(\gamma)d_{2,-2}^l(\gamma)
    \,, \label{23041202} \\
    \tilde{C}_l^{\Theta E}+\iu\tilde{C}_l^{\Theta B}
        &= 2\pi\sum_{l'mm'}\INT{}{(\cos\gamma)}{}{-1}{1}d^{l'}_{mm'}(\gamma)
        \left( C_{l'}^{\Theta E}+\iu C_{l'}^{\Theta B}\right)A_{l'}^X(\gamma)d_{20}^l(\gamma)
    \,. \label{23041203}
\end{align}
\end{widetext}
These results match those derived in Ref.~\cite{Gubitosi:2014cua}. Now, we have three equations for five unknown components, $\tilde{C}_l^{EE}$, $\tilde{C}_l^{BB}$, $\tilde{C}_l^{EB}$, $\tilde{C}_l^{\Theta E}$, and $\tilde{C}_l^{\Theta B}$. When the parity symmetry of the universe is conserved, the parity-odd components, $C_l^{EB}$, $C_l^{\Theta B}$, $\tilde{C}_l^{EB}$, and $\tilde{C}_l^{\Theta B}$, vanishes. Then, we have three equations for three unknown components and obtain the well-known results \cite{Challinor:2005jy}:
\begin{align}
    \tilde{C}_l^{EE}+\tilde{C}_l^{BB} &=
        2\pi\sum_{l'mm'}\INT{}{(\cos\gamma)}{}{-1}{1}d^{l'}_{mm'}(\gamma)
        \notag \\
        &\quad\times
        \left( C_{l'}^{EE}+C_{l'}^{BB}\right)A_{l'}^+(\gamma)d_{22}^l(\gamma)
    \,, \label{23041305}
\end{align}
\begin{align}
    \tilde{C}_l^{EE}-\tilde{C}_l^{BB} &=
        2\pi\sum_{l'mm'}\INT{}{(\cos\gamma)}{}{-1}{1}d^{l'}_{mm'}(\gamma)
        \notag \\
        &\quad\times
        \left( C_{l'}^{EE}-C_{l'}^{BB}\right)
        A_{l'}^-(\gamma)d_{2,-2}^l(\gamma)
    \,, \label{23041303}
\end{align}
\begin{align}
    \tilde{C}_l^{\Theta E} &= 2\pi\sum_{l'mm'}\INT{}{(\cos\gamma)}{}{-1}{1}d^{l'}_{mm'}(\gamma)
        \notag \\
        &\quad\times 
        C_{l'}^{\Theta E}A_{l'}^X(\gamma)d_{20}^l(\gamma)
    \,. \label{23041304}
\end{align}
When parity symmetry is violated by, e.g., cosmic birefringence, the parity-odd components do not vanish. In this case, taking the imaginary component of Eqs.~(\ref{23041201})-(\ref{23041203}), we obtain the expression for the lensed parity-odd power spectra: 
\begin{align}
    \tilde{C}_l^{EB}
        &= 2\pi\sum_{l'mm'}\INT{}{(\cos\gamma)}{}{-1}{1} d^{l'}_{mm'}(\gamma)
        \notag \\
        &\qquad \times
        C_{l'}^{EB} A_{l'}^-(\gamma) d_{2,-2}^l(\gamma)
    \,, \label{23041301} \\
    \tilde{C}_l^{\Theta B}
        &= 2\pi\sum_{l'mm'}\INT{}{(\cos\gamma)}{}{-1}{1} d^{l'}_{mm'}(\gamma)
        \notag \\
        &\qquad \times
        C_{l'}^{\Theta B}A_{l'}^X(\gamma)d_{20}^l(\gamma)
    \,. \label{23041302}
\end{align}
We find that Eq.~(\ref{23041301}) and Eq.~(\ref{23041302}) are almost the same as Eq.~(\ref{23041303}) and Eq.(\ref{23041304}), respectively. This means that the lensing effect on $C_l^{EB}$ and $C_l^{\Theta B}$ are the same as that on $C_l^{EE}-C_l^{BB}$ and $C_l^{\Theta E}$, respectively. These lensing corrections can be applied to $C_l^{EB}$ or $C_l^{\Theta B}$ induced by ALPs because the correction is commutable to the calculation of the Boltzmann equation, including the ALPs-induced cosmic birefringence\cite{Namikawa:2021:mode,Gubitosi:2014cua}. 

Since the parity-even and parity-odd components are completely separated into the real and imaginary parts, respectively, the lensing effect does not mix the parity-even and parity-odd components, i.e., the lensing conserves the parity symmetry. This fact is consistent with the result for the flat-sky $\Theta B$ power spectrum \cite{Saito:2007kt}. Therefore, gravitational lensing alone cannot explain the measured $EB$ power spectrum from the WMAP and Planck data \cite{Minami:2020odp,Diego-Palazuelos:2022dsq,Eskilt:2022cff}.

\section{Lensed power spectrum of cosmic birefringence}
\label{23042520}

We now present the numerical results of the ALPs-induced $EB$ power spectrum with the lensing correction. We also show the importance of the lensing effect on the ALPs parameter search in future experiments. We do not show the results for the $\Theta B$ power spectrum since its signal-to-noise is much lower than that of the $EB$ power spectrum. We choose the same cosmological parameters adopted in Ref.~\cite{Nakatsuka:2022epj}. 

\texttt{CLASS} implements the lensing correction to only the parity-even power spectra following Eqs.~(\ref{23041305})-(\ref{23041304}) and is not able to compute the lensed parity-odd power spectra. 
Thus, we include the lensing corrections, Eqs.~(\ref{23041301}) and (\ref{23041302}), to the parity-odd spectra in {\tt birefCLASS}. 

\begin{figure}[t]
  \centering
  \includegraphics[width=9cm]{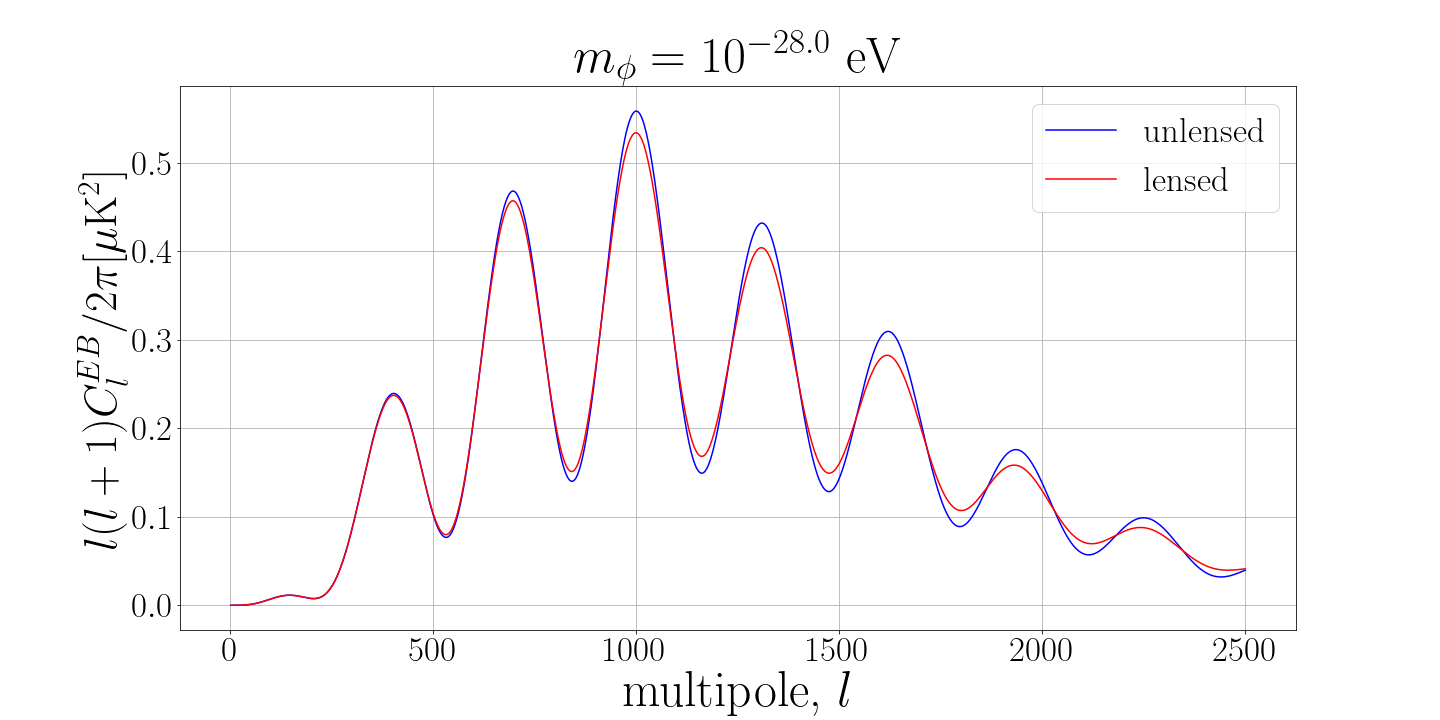}
  \caption{Unlensed (blue) and lensed (red) $EB$ power spectra induced by the ALPs with a mass of $m_\phi=10^{-28}~\mathrm{eV}$.}
  \label{fig:23042801}
\end{figure}

Fig.~\ref{fig:23042801} shows the unlensed and lensed $EB$ power spectra in the blue and red lines for $m_\phi=10^{-28}~\mathrm{eV}$. All the angular power spectra are multiplied by $T_0^2$, where $T_0=2.7255\times 10^{6}\,\mu$K is the CMB mean temperature. The amplitude of the $EB$ power spectrum scales with $A_{EB}\equiv g\phi_{\rm ini}/2$ where $\phi_{\rm ini}$ is an initial field value \cite{Nakatsuka:2022epj}. We choose $A_{EB}$ so that the $EB$ power spectrum reproduces the effective birefringence angle of $\beta=0.34\,$deg as defined in Eq.~(18) of Ref.~\cite{Nakatsuka:2022epj}.
\footnote{Ref.~\cite{Nakatsuka:2022epj} used $\beta=0.35^,$deg. In this paper, we choose a slightly different value; $\beta=0.34^,$deg following the latest value by Ref.~\cite{Eskilt:2022cff}.}
The lensing effect smears the acoustic peaks at small angular scales (high-$l$), while the changes at large-angular scales (low-$l$) are negligible. The typical displacement by lensing is a few arcminutes, and lensing mainly distorts the small-scale structure of CMB fluctuations. We also check other mass cases and find that the lensing effect generally mixes different multipoles and smears the acoustic peaks in the power spectrum. 

Next, to see the importance of the lensing effect on $EB$ power spectrum in future observations, we compare the difference between $C_l^{EB}$ and $\tilde{C}_l^{EB}$ with the observational errors for the LiteBIRD-like, SO-like, and S4-like experimental configurations. We compute the observational error on the $EB$ power spectrum per multipole as
\begin{align}
    (\sigma^{EB}_l)^2 = \frac{\tilde{C}_l^{EE,{\rm obs}}\tilde{C}_l^{BB,{\rm obs}}+\left(\tilde{C}_l^{EB}\right)^2}{f_{\mathrm{sky}}(2l+1)} 
    \,, \label{Eq:sigmaEB}
\end{align}
where $\tilde{C}_l^{EE,{\rm obs}}$ and $\tilde{C}_l^{BB,{\rm obs}}$ are the observed $E$- and $B$-mode power spectra, respectively. $f_\mathrm{sky}$ is a fraction of the sky which each experiment is supposed to observe. 
The observed power spectra are calculated as a sum of the lensed signal, $\tilde{C}_l^{XX}$, and noise power spectra, $N_l$. We assume a simple white noise spectrum with a beam-deconvolution effect for each experiment (e.g., \cite{Katayama:2011eh}):
\begin{equation}
\label{23042103}
    N_l \equiv \left(\frac{\sigma_p}{T_0}\frac{\pi}{10800}\right)^2\exp\left[\frac{l(l+1)}{8\ln2}\left(\theta\times\frac{\pi}{10800}\right)^2\right]
\end{equation}
where $\sigma_p$ is the noise level in the polarization map in the unit of $\mu$K-arcmin, and $\theta$ is the FWHM of a Gaussian beam in the unit of arcmin. 
We summarize these parameters in Table \ref{23042101} for each experiment. 
For LiteBIRD, we only consider the low-$l$ region because of the limited angular resolution. For SO and S4, we focus on the high-$l$ region because the large-angular scales are contaminated by atmospheric noise in ground-based observations. The minimum and maximum multipole of the power spectrum used in our analysis, $l_\mathrm{min}$ and $l_\mathrm{max}$, are also summarized in Table \ref{23042101}. 

\begin{table}[t]
\begin{tabular}{cccc}
\hline
                   & LiteBIRD & SO & S4 \\ \hline
$\sigma_p$ [$\mu$K-arcmin]  & 2        & 6                  & 1      \\
$\theta$ [arcmin]  & 30       & 1                  & 1      \\ \hline
$f_{\mathrm{sky}}$ & 0.7      & 0.4                & 0.4    \\
$l_{\mathrm{min}}$ & 2        & 100                & 100    \\
$l_{\mathrm{max}}$ & 500      & 2500               & 2500   \\ \hline
\end{tabular}
\caption{The experimental configuration for future experiments we consider. The ``SO'' and ``S4'' correspond to the wide-field large-aperture telescopes of the Simons Observatory and CMB-S4, respectively.}\label{23042101}
\end{table}

\begin{figure}[t]
  \centering
  \includegraphics[width=8.5cm]{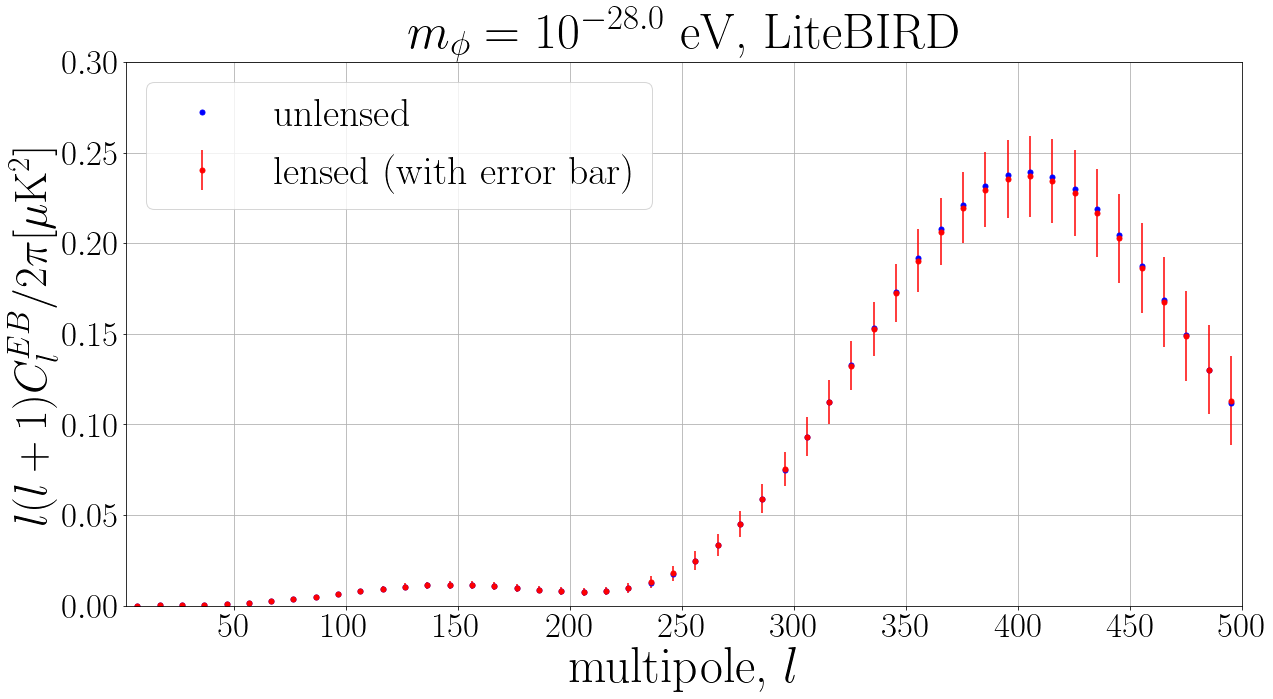}
  \caption{Comparison of the lensing effect with the observational error bars for LiteBIRD. The blue and red points denote the unlensed and lensed power spectra with the multipole binning, respectively. Only the low-$l$ region is shown.}
  \label{fig:23042810}
\end{figure}

\begin{figure}[t]
  \centering
  \includegraphics[width=8.5cm]{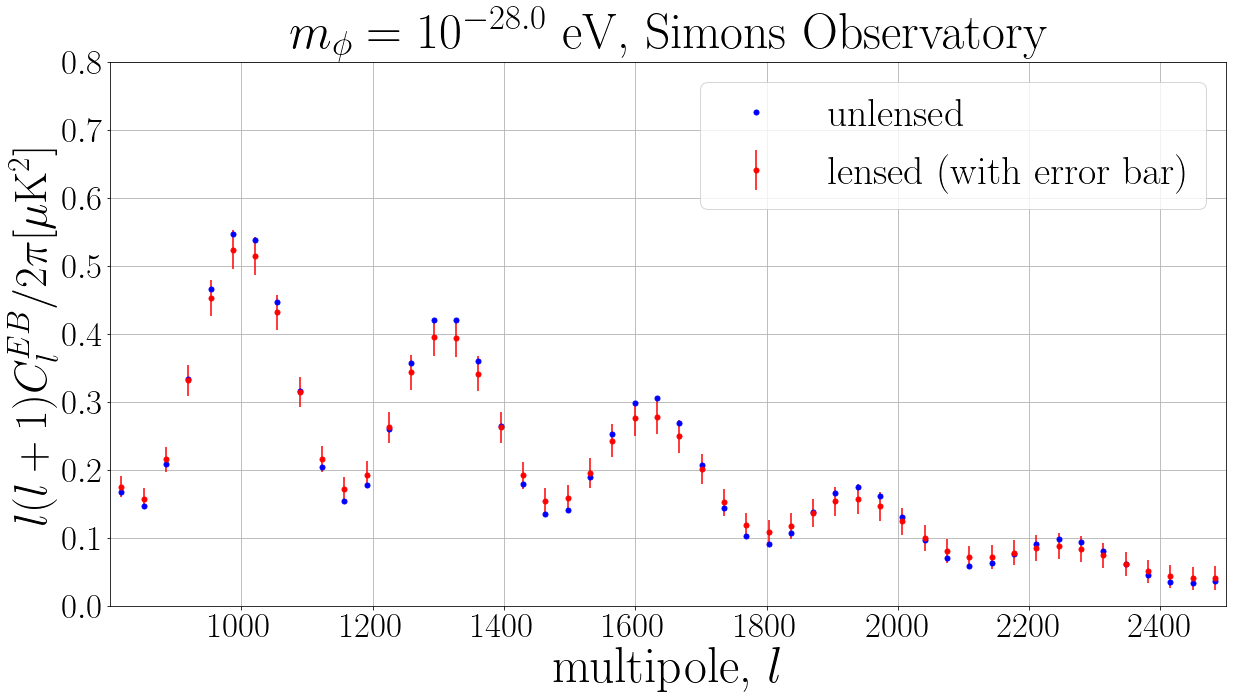}
  \caption{Same as Fig.~\ref{fig:23042810} but for SO. Only the high-$l$ region is shown.}
  \label{fig:23042811}
\end{figure}

\begin{figure}[t]
  \centering
  \includegraphics[width=8.5cm]{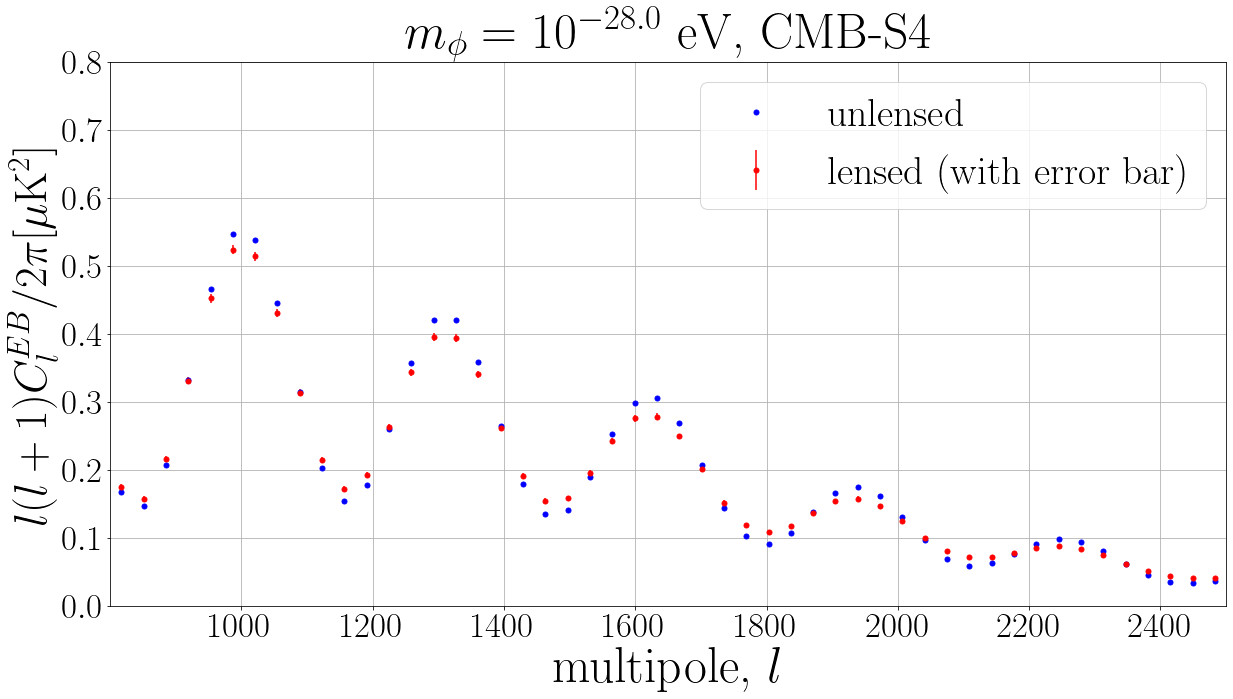}
  \caption{Same as Fig.~\ref{fig:23042811} but for S4.}
  \label{fig:23042812}
\end{figure}

Figs.~\ref{fig:23042810}-\ref{fig:23042812} show the unlensed (blue points) and lensed (red points) power spectra with the observational error bars for each experiment when $m_\phi=10^{-28}~\mathrm{eV}$. 
For each experiment, the multiple range given in Table \ref{23042101} is divided into $50$ equally-sized bins.
The red error bars show the expected statistical uncertainty of the $EB$ power spectrum for each experiment. 
The lensing effect on the $EB$ power spectrum is negligible at low-$l$, and the difference between $C_l^{EB}$ and $\tilde{C}_l^{EB}$ is much smaller than the error bars for LiteBIRD. On the other hand, the difference is comparable to the error bars for SO and is much larger than the error bars for S4. Thus, we must include the lensing effect on the $EB$ power spectrum in future ground-based observations to probe cosmic birefringence. 

\section{Lensing effect on ALPs search}
\label{23042521}

In this section, we perform a more quantitative analysis to show the importance of the lensing effect on probing cosmic birefringence. 

We first show the impact on the ALPs parameter search with a similar forecast setup to Ref.~\cite{Nakatsuka:2022epj}. 
They considered a set of three free parameters for the ALPs search, $\bm{p}=(m_\phi, A\equiv A_{EB}/A_{EB}^{0.34^\circ}, \alpha)$. Here, $m_\phi$ is the mass of ALPs, $A_{EB}^{0.34^\circ}$ is the value of $g\phi_{\rm ini}/2$ which reproduces the effective birefringence angle of $0.34\,$deg, and $\alpha$ is the instrumental miscalibration angle. For a given mock data of the $EB$ spectrum computed with fiducial parameters, $C_l^{EB,{\rm fid}}$, they evaluated the following chi-square for each $\bm{p}$\footnote{This form for $\chi^2$ is motivated by the fact that the cross-power spectrum of the harmonic coefficients is distributed as a Gaussian \cite{Gupta&Nagar:1999,Hamimeche:2008ai}.}: 
\begin{equation}
  \label{23010352}
  \begin{aligned}
    &\chi^2(\bm{p})
    \equiv \sum_{l=l_{\mathrm{min}}}^{l_{\mathrm{max}}}
    \frac{\left[C_l^{EB,\mathrm{fid}}-C_l^{EB}(\bm{p})\right]^2}
    {(\bar{\sigma}^{EB}_l)^2}
  \,,
  \end{aligned}
\end{equation}
where $\bar{\sigma}_l^{EB}$ is the error of Eq.~\eqref{Eq:sigmaEB} but without $(\tilde{C}_l^{EB})^2$ in the numerator. 
The error contour is then obtained by computing the following posterior distribution: 
\begin{equation}
    {\cal P}_\mathrm{post}(\bm{p}) \propto \exp\left[-\frac{1}{2}\chi^2(\bm{p})\right] \exp
    \left(-\frac{\alpha^2}{2\sigma_\alpha^2}\right)
    \,. 
\end{equation}
Here, the second exponential function is a prior distribution for $\alpha$, and we choose $\sigma_\alpha=0.1\,$deg for our baseline calculation, which the future CMB experiments would achieve \cite{Johnson_2015,Nati:2017lnn,Casas}. 

We repeat the same calculation above but with the lensed $EB$ power spectrum for $\chi^2$:
\begin{equation}
  \label{23010351}
    \chi^2(\bm{p})
    = \sum_{l=l_{\mathrm{min}}}^{l_{\mathrm{max}}}
    \frac{\left[ \tilde{C}_l^{EB,\mathrm{fid}}-\tilde{C}_l^{EB}(\bm{p}) \right]^2}
    {(\sigma^{EB}_l)^2}
  \,, 
\end{equation}
where $\tilde{C}_l^{EB,\mathrm{fid}}$ denotes the mock data with lensing. We compute the posterior distribution of Eq.~(\ref{23010351}) with a set of fiducial parameters, $m_\phi=10^{-28}\,$eV, $A=1$, and $\alpha=0\,$deg. We find that including the lensing changes the error contour only slightly. This means that the parameter constraints mostly come from the statistical uncertainty, $\sigma_l^{EB}$. Our result shows that the forecast by Ref.~\cite{Nakatsuka:2022epj}, where the unlensed power spectrum is used, would be close to those obtained with the lensing corrections. 

\begin{figure}[t]
  \centering
  \includegraphics[width=8cm]{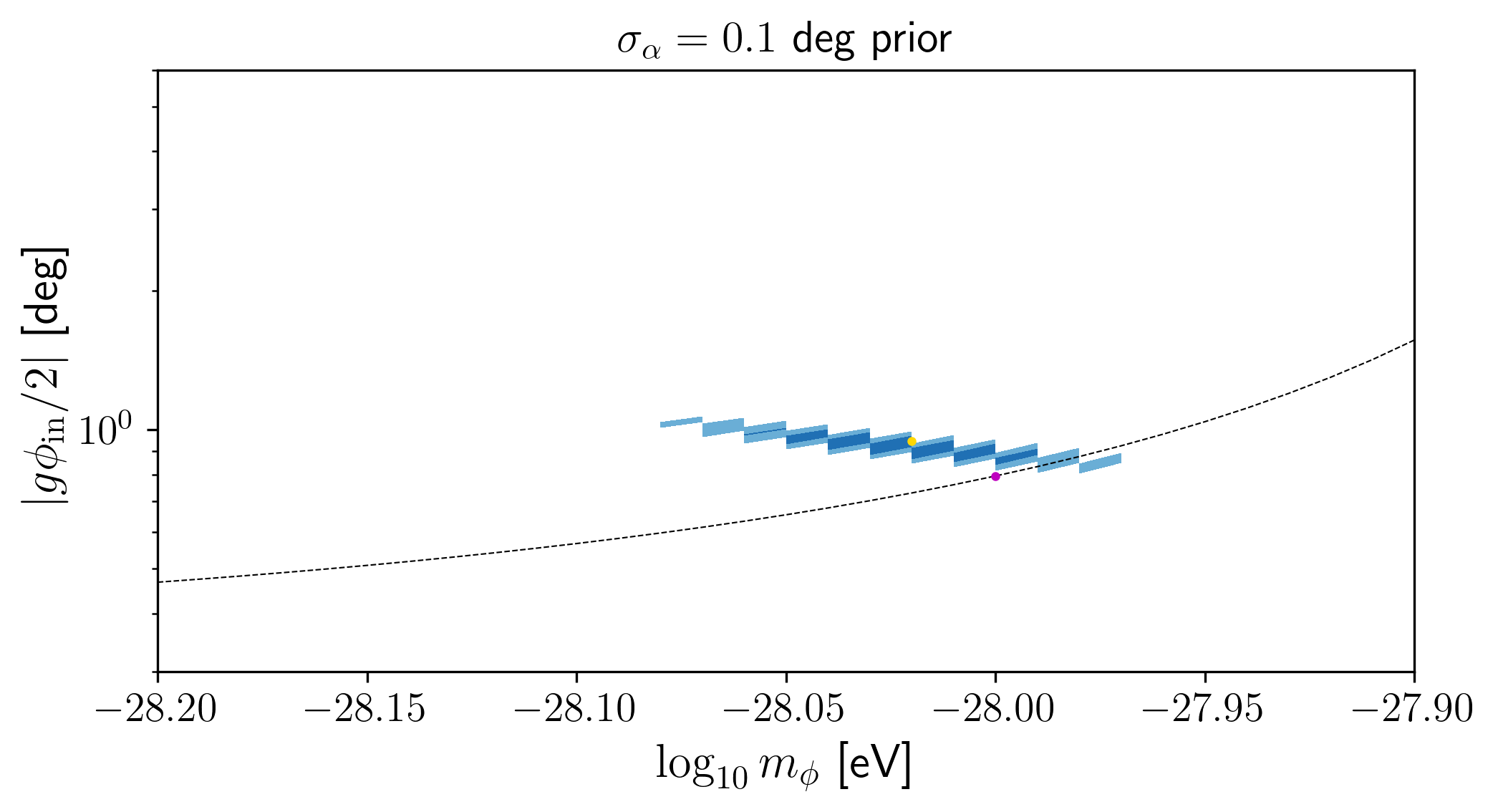}
  \caption{Error contours of the ALPs parameters estimated with the {\it unlensed} power spectrum for a theoretical prediction. The pale and dark blue areas show $1\,\sigma$ and $2\,\sigma$ regions, respectively. We marginalize over $\alpha$. The red dot corresponds to the fiducial value. The yellow dot is the value that minimizes Eq.~(\ref{23042501}). The dashed line shows the values of $|g\phi_\mathrm{ini}/2|$ and $m_\phi$, which roughly explain $\beta=0.34\,$deg when regarding it as a constant rotation.}
  \label{fig:23042507}
\end{figure}

However, in practice, we need the lensing correction to the $EB$ power spectrum for the ALPs search because observed data are lensed. 
Thus, we test how well the unlensed $EB$ spectrum fits the lensed $EB$ spectrum. To quantify this, let us introduce
\begin{align}
    &\hat{\chi}^2_{\mathcal{A}}(\bm{p})
    \equiv
    \sum_{l=l_{\mathrm{min}}}^{l_{\mathrm{max}}}
    \notag \\
    &\times \frac{ \left[ \tilde{C}_l^{EB,\mathrm{fid}}
    - \left\{ \mathcal{A} \tilde{C}_l^{EB}(\bm{p}) + (1-\mathcal{A})C_l^{EB}(\bm{p}) \right\}
    \right]^2}
    {(\sigma_l^{EB})^2}
    \,, \label{23042502}
\end{align}
where $\mathcal{A}$ is a parameter expressing the degree of lensing; the fitting model is lensed (unlensed) if $\mathcal{A}=1$ ($\mathcal{A}=0$). 
We compute the minimum value of $\hat{\chi}^2_{\mathcal{A}}(\bm{p})$ for $\mathcal{A}=0$ and obtain\footnote{The second term, $\min_{\bm{p},\mathcal{A}}\hat{\chi}^2_{\mathcal{A}}(\bm{p})$, is zero by definition.}: 
\begin{equation}
    \Delta\chi^2 = \min_{\bm{p}}\hat{\chi}^2_{\mathcal{A}=0}(\bm{p})-\min_{\bm{p},\mathcal{A}}\hat{\chi}^2_{\mathcal{A}}(\bm{p})
    \,. 
\end{equation}
We then derive a $p$-value of $\Delta\chi^2$ assuming that $\Delta\chi^2$ obeys a $\chi^2$ distribution with $1$ degree of freedom. This analysis is motivated by the fact that $\Delta\chi^2$ is considered as a likelihood ratio for two models where the null model fixes $\mathcal{A}=0$ with three parameters, $\bm{p}$, and the alternate model varies $\mathcal{A}$ with four parameters, $\bm{p}$ and $\mathcal{A}$. For such a nested case, $\Delta\chi^2$ would approximately obey the $\chi^2$ distribution where its degree of freedom is equivalent to the difference in the number of free parameters between the two models. 
We find that the $p$-value from the above $\Delta\chi^2$ is $0.76$ for LiteBIRD. On the other hand $\Delta\chi^2=30$ and $953$ for SO and S4, respectively, and the corresponding $p$-values are close to zero. The deviation by the lensing effect is not significant for LiteBIRD. On the other hand, for SO and S4, the unlensed power spectrum does not fit an observed (lensed) power spectrum. This means we should not use the unlensed power spectrum to analyze SO and S4 data.

The estimated parameters are biased if we force ourselves to use the unlensed power spectrum. 
To quantify the bias caused by the use of the unlensed power spectrum, we prepared the lensed $EB$ power spectra as mock data and fit this data with the unlensed $EB$ power spectrum. Specifically, we compute
\begin{equation}
    \label{23042501}
    \chi^2(\bm{p})=
    \sum_{l=l_{\mathrm{min}}}^{l_{\mathrm{max}}} 
    \frac{\left[\tilde{C}_l^{EB,\mathrm{fid}}-C_l^{EB}(\bm{p}) \right]^2}
    {(\sigma_l^{EB})^2}
    \,. 
\end{equation}
We assume $m_\phi=10^{-28}~\mathrm{eV}$, $A=1$, and $\alpha=0\,$deg for the mock data. Since the lensing effect is unimportant on large-angular scales, the bias is negligible compared to the error contour for LiteBIRD. In the SO case, although the unlensed power spectrum does not fit the lensed power spectrum, the bias is not significant compared to the $2\,\sigma$ error contour.
For S4, we show the result in Fig.~\ref{fig:23042507}. The shaded pale and dark blue regions indicate the $1\,\sigma$ and $2\,\sigma$ uncertainties. The red point is the fiducial value, and the yellow point denotes the points where $\chi^2$ of Eq.~(\ref{23042501}) is minimized. The fiducial values are outside the error contour. We also check the dependence on the prior on $\alpha$ by increasing $\sigma_\alpha$ to $0.5\,$deg. The size of the error contour only increases along the degeneracy direction, and the true value is still well outside of the $2\,\sigma$ error contour. The estimated parameters are substantially biased when we use more sensitive data such as S4 to analyze cosmic birefringence.

\section{Summary and Discussion} \label{sec:summary}
We explored the effect of lensing on the ALPs-induced parity-odd power spectra by improving \texttt{birefCLASS} \cite{Nakatsuka:2022epj}. 
We showed that the lensing effect smears the peaks in the $EB$ power spectrum at high-$l$. We found that the modification by the lensing effect is much larger than the observational errors for SO and S4 at small angular scales. In contrast, the lensing effect is negligible for LiteBIRD. We also explored how the ignorance of the lensing effect in the parameter search leads to a bias and found that the bias is not negligible compared to the error bar for S4. For SO and S4, the lensing effect should be included to fit the observed $EB$ power spectrum. We conclude that we need an accurate theoretical prediction incorporating the lensing correction for ALPs searches with cosmic birefringence in future high-$l$ CMB observations.


We have shown that the lensing smears the peaks of $C_l^{EB}$ at high-$l$. This modification to $C_l^{EB}$ is unlikely to be produced by other physical effects, including cosmic birefringence and changes in cosmological parameters. On the other hand, instrumental systematics could modify the $EB$ power spectrum, which mimics the lensing effect. For example, the pointing error also distorts the polarization map in the same way as lensing. However, the typical pointing error is expected to be at a few arcseconds for future CMB observations (e.g., \cite{Mirmelstein:2020,Nagata:2021}). Since the typical lensing displacement is at a few arcminutes, the impact of the pointing error on the $EB$ power spectrum would be much smaller than that of lensing. 

The beam systematics also lead to a modification in the small-scale CMB anisotropies where lensing is important. A measured $EB$ power spectrum at high-$l$ could have a bias due to imperfect knowledge of the beam. However, the beam systematics also affect all the other CMB power spectra. The temperature and $E$-mode power spectra have much larger signal-to-noise than the $EB$ power spectrum. Thus, the beam uncertainties are constrained by the temperature and $E$-mode power spectra so that their impact on the $EB$ power spectrum is negligible. 

In this work, we focused on the lensing correction to $C_l^{EB}$ and $C_l^{\Theta B}$ in the context of ALPs. The lensing correction to the parity-odd power spectra does not depend on the model of the unlensed $EB$ power spectrum and is also important for cosmic birefringence sourced from other physical origins, including primordial cases, e.g., \cite{Saito:2007kt,Thorne:2017jft,Fujita:2022qlk,Gonzalez:2022mcx,Gasparotto:2022uqo,Obata:2021nql}. 
Note that Refs.~\cite{Murai:2022:EDE,Eskilt:2023:EDE} recently studied the cosmic birefringence from the early dark energy model where the lensing effect on the $EB$ power spectrum was included using our code. 
When writing this paper, Ref.~\cite{Yin:2023} explores cosmic birefringence induced by the $\alpha$-attractor and the Rock`n' Roll field models where the lensing correction is adopted. The lensing correction to the parity-odd power spectra is necessary for studying cosmic birefringence at high-$l$. 

In the coming few years, BICEP\cite{Moncelsi:2020ppj}, Simons Array\cite{POLARBEAR:2015ixw}, and ACT\cite{2021ITAS...3163334L} will analyze new observational data containing CMB polarization. Looking further into the future, LiteBIRD \cite{LiteBIRD:2022:PTEP}, SO \cite{SimonsObservatory}, and S4 \cite{CMBS4:r-forecast} will provide profoundly sophisticated data. The code developed in this paper will be useful to constrain cosmic birefringence from these data.

\begin{acknowledgments}
We thank Eiichiro Komatsu for his insightful discussion on this work, and Jun'ichi Yokoyama and Kohei Kamada for their guidance, support, and fruitful discussion. We also thank Akito Kusaka, Tomotake Matsumura, Kai Murai, Ippei Obata, and Maresuke Shiraishi for their comments and supports, and the participants of the workshop YITP-T-21-08 on “Upcoming CMB Observation and Cosmology,” held at the Yukawa Institute for Theoretical Physics, where this work was initiated. This work was supported in part by the Forefront Physics and Mathematics Program to Drive Transformation (FoPM), a World-leading Innovative Graduate Study (WINGS) Program, the University of Tokyo (FN), and
JSPS KAKENHI Grant No. JP20H05859, and No. JP22K03682 (TN). 
The Kavli IPMU is supported by World Premier International Research Center Initiative (WPI Initiative), MEXT, Japan.
\end{acknowledgments}

\bibliographystyle{apsrev4-2}
\bibliography{apssamp_v2}

\providecommand{\noopsort}[1]{}\providecommand{\singleletter}[1]{#1}%
\begin{thebibliography}{67}%
\makeatletter
\providecommand \@ifxundefined [1]{%
 \@ifx{#1\undefined}
}%
\providecommand \@ifnum [1]{%
 \ifnum #1\expandafter \@firstoftwo
 \else \expandafter \@secondoftwo
 \fi
}%
\providecommand \@ifx [1]{%
 \ifx #1\expandafter \@firstoftwo
 \else \expandafter \@secondoftwo
 \fi
}%
\providecommand \natexlab [1]{#1}%
\providecommand \enquote  [1]{``#1''}%
\providecommand \bibnamefont  [1]{#1}%
\providecommand \bibfnamefont [1]{#1}%
\providecommand \citenamefont [1]{#1}%
\providecommand \href@noop [0]{\@secondoftwo}%
\providecommand \href [0]{\begingroup \@sanitize@url \@href}%
\providecommand \@href[1]{\@@startlink{#1}\@@href}%
\providecommand \@@href[1]{\endgroup#1\@@endlink}%
\providecommand \@sanitize@url [0]{\catcode `\\12\catcode `\$12\catcode
  `\&12\catcode `\#12\catcode `\^12\catcode `\_12\catcode `\%12\relax}%
\providecommand \@@startlink[1]{}%
\providecommand \@@endlink[0]{}%
\providecommand \url  [0]{\begingroup\@sanitize@url \@url }%
\providecommand \@url [1]{\endgroup\@href {#1}{\urlprefix }}%
\providecommand \urlprefix  [0]{URL }%
\providecommand \Eprint [0]{\href }%
\providecommand \doibase [0]{https://doi.org/}%
\providecommand \selectlanguage [0]{\@gobble}%
\providecommand \bibinfo  [0]{\@secondoftwo}%
\providecommand \bibfield  [0]{\@secondoftwo}%
\providecommand \translation [1]{[#1]}%
\providecommand \BibitemOpen [0]{}%
\providecommand \bibitemStop [0]{}%
\providecommand \bibitemNoStop [0]{.\EOS\space}%
\providecommand \EOS [0]{\spacefactor3000\relax}%
\providecommand \BibitemShut  [1]{\csname bibitem#1\endcsname}%
\let\auto@bib@innerbib\@empty
\bibitem [{\citenamefont {Komatsu}(2022)}]{Komatsu:2022:review}%
  \BibitemOpen
  \bibfield  {author} {\bibinfo {author} {\bibfnamefont {E.}~\bibnamefont
  {Komatsu}},\ }\href {https://doi.org/10.1038/s42254-022-00452-4} {\bibfield
  {journal} {\bibinfo  {journal} {Nature Rev. Phys.}\ }\textbf {\bibinfo
  {volume} {4}},\ \bibinfo {pages} {452} (\bibinfo {year} {2022})},\ \Eprint
  {https://arxiv.org/abs/2202.13919} {arXiv:2202.13919 [astro-ph.CO]}
  \BibitemShut {NoStop}%
\bibitem [{\citenamefont {Carroll}(1998)}]{Carroll:1998zi}%
  \BibitemOpen
  \bibfield  {author} {\bibinfo {author} {\bibfnamefont {S.~M.}\ \bibnamefont
  {Carroll}},\ }\href {https://doi.org/10.1103/PhysRevLett.81.3067} {\bibfield
  {journal} {\bibinfo  {journal} {\prl}\ }\textbf {\bibinfo {volume} {81}},\
  \bibinfo {pages} {3067} (\bibinfo {year} {1998})},\ \Eprint
  {https://arxiv.org/abs/astro-ph/9806099} {arXiv:astro-ph/9806099}
  \BibitemShut {NoStop}%
\bibitem [{\citenamefont {Carroll}\ \emph {et~al.}(1990)\citenamefont
  {Carroll}, \citenamefont {Field},\ and\ \citenamefont
  {Jackiw}}]{Carroll:1989vb}%
  \BibitemOpen
  \bibfield  {author} {\bibinfo {author} {\bibfnamefont {S.~M.}\ \bibnamefont
  {Carroll}}, \bibinfo {author} {\bibfnamefont {G.~B.}\ \bibnamefont {Field}},\
  and\ \bibinfo {author} {\bibfnamefont {R.}~\bibnamefont {Jackiw}},\ }\href
  {https://doi.org/10.1103/PhysRevD.41.1231} {\bibfield  {journal} {\bibinfo
  {journal} {\prd}\ }\textbf {\bibinfo {volume} {41}},\ \bibinfo {pages} {1231}
  (\bibinfo {year} {1990})}\BibitemShut {NoStop}%
\bibitem [{\citenamefont {Carroll}\ and\ \citenamefont
  {Field}(1991)}]{Carroll:1991zs}%
  \BibitemOpen
  \bibfield  {author} {\bibinfo {author} {\bibfnamefont {S.~M.}\ \bibnamefont
  {Carroll}}\ and\ \bibinfo {author} {\bibfnamefont {G.~B.}\ \bibnamefont
  {Field}},\ }\href {https://doi.org/10.1103/PhysRevD.43.3789} {\bibfield
  {journal} {\bibinfo  {journal} {\prd}\ }\textbf {\bibinfo {volume} {43}},\
  \bibinfo {pages} {3789} (\bibinfo {year} {1991})}\BibitemShut {NoStop}%
\bibitem [{\citenamefont {Harari}\ and\ \citenamefont
  {Sikivie}(1992)}]{Harari:1992ea}%
  \BibitemOpen
  \bibfield  {author} {\bibinfo {author} {\bibfnamefont {D.}~\bibnamefont
  {Harari}}\ and\ \bibinfo {author} {\bibfnamefont {P.}~\bibnamefont
  {Sikivie}},\ }\href {https://doi.org/10.1016/0370-2693(92)91363-E} {\bibfield
   {journal} {\bibinfo  {journal} {\plb}\ }\textbf {\bibinfo {volume} {289}},\
  \bibinfo {pages} {67} (\bibinfo {year} {1992})}\BibitemShut {NoStop}%
\bibitem [{\citenamefont {Lue}\ \emph {et~al.}(1999)\citenamefont {Lue},
  \citenamefont {Wang},\ and\ \citenamefont {Kamionkowski}}]{Lue:1998mq}%
  \BibitemOpen
  \bibfield  {author} {\bibinfo {author} {\bibfnamefont {A.}~\bibnamefont
  {Lue}}, \bibinfo {author} {\bibfnamefont {L.-M.}\ \bibnamefont {Wang}},\ and\
  \bibinfo {author} {\bibfnamefont {M.}~\bibnamefont {Kamionkowski}},\ }\href
  {https://doi.org/10.1103/PhysRevLett.83.1506} {\bibfield  {journal} {\bibinfo
   {journal} {Phys. Rev. Lett.}\ }\textbf {\bibinfo {volume} {83}},\ \bibinfo
  {pages} {1506} (\bibinfo {year} {1999})},\ \Eprint
  {https://arxiv.org/abs/astro-ph/9812088} {arXiv:astro-ph/9812088}
  \BibitemShut {NoStop}%
\bibitem [{\citenamefont {Feng}\ \emph {et~al.}(2005)\citenamefont {Feng},
  \citenamefont {Li}, \citenamefont {Li},\ and\ \citenamefont
  {Zhang}}]{Feng:2004mq}%
  \BibitemOpen
  \bibfield  {author} {\bibinfo {author} {\bibfnamefont {B.}~\bibnamefont
  {Feng}}, \bibinfo {author} {\bibfnamefont {H.}~\bibnamefont {Li}}, \bibinfo
  {author} {\bibfnamefont {M.-z.}\ \bibnamefont {Li}},\ and\ \bibinfo {author}
  {\bibfnamefont {X.-m.}\ \bibnamefont {Zhang}},\ }\href
  {https://doi.org/10.1016/j.physletb.2005.06.009} {\bibfield  {journal}
  {\bibinfo  {journal} {Phys. Lett. B}\ }\textbf {\bibinfo {volume} {620}},\
  \bibinfo {pages} {27} (\bibinfo {year} {2005})},\ \Eprint
  {https://arxiv.org/abs/hep-ph/0406269} {arXiv:hep-ph/0406269} \BibitemShut
  {NoStop}%
\bibitem [{\citenamefont {Fujita}\ \emph {et~al.}(2021)\citenamefont {Fujita},
  \citenamefont {Murai}, \citenamefont {Nakatsuka},\ and\ \citenamefont
  {Tsujikawa}}]{Fujita:2020ecn}%
  \BibitemOpen
  \bibfield  {author} {\bibinfo {author} {\bibfnamefont {T.}~\bibnamefont
  {Fujita}}, \bibinfo {author} {\bibfnamefont {K.}~\bibnamefont {Murai}},
  \bibinfo {author} {\bibfnamefont {H.}~\bibnamefont {Nakatsuka}},\ and\
  \bibinfo {author} {\bibfnamefont {S.}~\bibnamefont {Tsujikawa}},\ }\href
  {https://doi.org/10.1103/PhysRevD.103.043509} {\bibfield  {journal} {\bibinfo
   {journal} {\prd}\ }\textbf {\bibinfo {volume} {103}},\ \bibinfo {pages}
  {043509} (\bibinfo {year} {2021})},\ \Eprint
  {https://arxiv.org/abs/2011.11894} {arXiv:2011.11894 [astro-ph.CO]}
  \BibitemShut {NoStop}%
\bibitem [{\citenamefont {Finelli}\ and\ \citenamefont
  {Galaverni}(2009)}]{Finelli:2008}%
  \BibitemOpen
  \bibfield  {author} {\bibinfo {author} {\bibfnamefont {F.}~\bibnamefont
  {Finelli}}\ and\ \bibinfo {author} {\bibfnamefont {M.}~\bibnamefont
  {Galaverni}},\ }\href {https://doi.org/10.1103/PhysRevD.79.063002} {\bibfield
   {journal} {\bibinfo  {journal} {\prd}\ }\textbf {\bibinfo {volume} {79}},\
  \bibinfo {pages} {063002} (\bibinfo {year} {2009})},\ \Eprint
  {https://arxiv.org/abs/0802.4210} {0802.4210} \BibitemShut {NoStop}%
\bibitem [{\citenamefont {Fedderke}\ \emph {et~al.}(2019)\citenamefont
  {Fedderke}, \citenamefont {Graham},\ and\ \citenamefont
  {Rajendran}}]{Fedderke:2019ajk}%
  \BibitemOpen
  \bibfield  {author} {\bibinfo {author} {\bibfnamefont {M.~A.}\ \bibnamefont
  {Fedderke}}, \bibinfo {author} {\bibfnamefont {P.~W.}\ \bibnamefont
  {Graham}},\ and\ \bibinfo {author} {\bibfnamefont {S.}~\bibnamefont
  {Rajendran}},\ }\href {https://doi.org/10.1103/PhysRevD.100.015040}
  {\bibfield  {journal} {\bibinfo  {journal} {\prd}\ }\textbf {\bibinfo
  {volume} {100}},\ \bibinfo {pages} {015040} (\bibinfo {year} {2019})},\
  \Eprint {https://arxiv.org/abs/1903.02666} {arXiv:1903.02666 [astro-ph.CO]}
  \BibitemShut {NoStop}%
\bibitem [{\citenamefont {Panda}\ \emph {et~al.}(2011)\citenamefont {Panda},
  \citenamefont {Sumitomo},\ and\ \citenamefont {Trivedi}}]{Panda:2010uq}%
  \BibitemOpen
  \bibfield  {author} {\bibinfo {author} {\bibfnamefont {S.}~\bibnamefont
  {Panda}}, \bibinfo {author} {\bibfnamefont {Y.}~\bibnamefont {Sumitomo}},\
  and\ \bibinfo {author} {\bibfnamefont {S.~P.}\ \bibnamefont {Trivedi}},\
  }\href {https://doi.org/10.1103/PhysRevD.83.083506} {\bibfield  {journal}
  {\bibinfo  {journal} {\prd}\ }\textbf {\bibinfo {volume} {83}},\ \bibinfo
  {pages} {083506} (\bibinfo {year} {2011})},\ \Eprint
  {https://arxiv.org/abs/1011.5877} {arXiv:1011.5877 [hep-th]} \BibitemShut
  {NoStop}%
\bibitem [{\citenamefont {Myers}\ and\ \citenamefont
  {Pospelov}(2003)}]{Myers:2003fd}%
  \BibitemOpen
  \bibfield  {author} {\bibinfo {author} {\bibfnamefont {R.~C.}\ \bibnamefont
  {Myers}}\ and\ \bibinfo {author} {\bibfnamefont {M.}~\bibnamefont
  {Pospelov}},\ }\href {https://doi.org/10.1103/PhysRevLett.90.211601}
  {\bibfield  {journal} {\bibinfo  {journal} {\prl}\ }\textbf {\bibinfo
  {volume} {90}},\ \bibinfo {pages} {211601} (\bibinfo {year} {2003})},\
  \Eprint {https://arxiv.org/abs/hep-ph/0301124} {arXiv:hep-ph/0301124}
  \BibitemShut {NoStop}%
\bibitem [{\citenamefont {Arvanitaki}\ \emph {et~al.}(2010)\citenamefont
  {Arvanitaki}, \citenamefont {Dimopoulos}, \citenamefont {Dubovsky},
  \citenamefont {Kaloper},\ and\ \citenamefont
  {March-Russell}}]{Arvanitaki:2009fg}%
  \BibitemOpen
  \bibfield  {author} {\bibinfo {author} {\bibfnamefont {A.}~\bibnamefont
  {Arvanitaki}}, \bibinfo {author} {\bibfnamefont {S.}~\bibnamefont
  {Dimopoulos}}, \bibinfo {author} {\bibfnamefont {S.}~\bibnamefont
  {Dubovsky}}, \bibinfo {author} {\bibfnamefont {N.}~\bibnamefont {Kaloper}},\
  and\ \bibinfo {author} {\bibfnamefont {J.}~\bibnamefont {March-Russell}},\
  }\href {https://doi.org/10.1103/PhysRevD.81.123530} {\bibfield  {journal}
  {\bibinfo  {journal} {\prd}\ }\textbf {\bibinfo {volume} {81}},\ \bibinfo
  {pages} {123530} (\bibinfo {year} {2010})},\ \Eprint
  {https://arxiv.org/abs/0905.4720} {arXiv:0905.4720 [hep-th]} \BibitemShut
  {NoStop}%
\bibitem [{\citenamefont {Minami}\ and\ \citenamefont
  {Komatsu}(2020)}]{Minami:2020odp}%
  \BibitemOpen
  \bibfield  {author} {\bibinfo {author} {\bibfnamefont {Y.}~\bibnamefont
  {Minami}}\ and\ \bibinfo {author} {\bibfnamefont {E.}~\bibnamefont
  {Komatsu}},\ }\href {https://doi.org/10.1103/PhysRevLett.125.221301}
  {\bibfield  {journal} {\bibinfo  {journal} {\prl}\ }\textbf {\bibinfo
  {volume} {125}},\ \bibinfo {pages} {221301} (\bibinfo {year} {2020})},\
  \Eprint {https://arxiv.org/abs/2011.11254} {arXiv:2011.11254 [astro-ph.CO]}
  \BibitemShut {NoStop}%
\bibitem [{\citenamefont {Diego-Palazuelos}\ \emph {et~al.}(2022)\citenamefont
  {Diego-Palazuelos} \emph {et~al.}}]{Diego-Palazuelos:2022dsq}%
  \BibitemOpen
  \bibfield  {author} {\bibinfo {author} {\bibfnamefont {P.}~\bibnamefont
  {Diego-Palazuelos}} \emph {et~al.},\ }\href
  {https://doi.org/10.1103/PhysRevLett.128.091302} {\bibfield  {journal}
  {\bibinfo  {journal} {\prl}\ }\textbf {\bibinfo {volume} {128}},\ \bibinfo
  {pages} {091302} (\bibinfo {year} {2022})},\ \Eprint
  {https://arxiv.org/abs/2201.07682} {arXiv:2201.07682 [astro-ph.CO]}
  \BibitemShut {NoStop}%
\bibitem [{\citenamefont {Eskilt}\ and\ \citenamefont
  {Komatsu}(2022)}]{Eskilt:2022cff}%
  \BibitemOpen
  \bibfield  {author} {\bibinfo {author} {\bibfnamefont {J.~R.}\ \bibnamefont
  {Eskilt}}\ and\ \bibinfo {author} {\bibfnamefont {E.}~\bibnamefont
  {Komatsu}},\ }\href {https://doi.org/10.1103/PhysRevD.106.063503} {\bibfield
  {journal} {\bibinfo  {journal} {\prd}\ }\textbf {\bibinfo {volume} {106}},\
  \bibinfo {pages} {063503} (\bibinfo {year} {2022})},\ \Eprint
  {https://arxiv.org/abs/2205.13962} {arXiv:2205.13962 [astro-ph.CO]}
  \BibitemShut {NoStop}%
\bibitem [{\citenamefont {{LiteBIRD
  Collaboration}}(2022)}]{LiteBIRD:2022:PTEP}%
  \BibitemOpen
  \bibfield  {author} {\bibinfo {author} {\bibnamefont {{LiteBIRD
  Collaboration}}},\ }\bibfield  {journal} {\bibinfo  {journal} {\ptep}\ }\href
  {https://doi.org/10.1093/ptep/ptac150} {10.1093/ptep/ptac150} (\bibinfo
  {year} {2022}),\ \Eprint {https://arxiv.org/abs/2202.02773} {arXiv:2202.02773
  [astro-ph.IM]} \BibitemShut {NoStop}%
\bibitem [{\citenamefont {{The Simons Observatory
  Collaboration}}(2019)}]{SimonsObservatory}%
  \BibitemOpen
  \bibfield  {author} {\bibinfo {author} {\bibnamefont {{The Simons Observatory
  Collaboration}}},\ }\href {https://doi.org/10.1088/1475-7516/2019/02/056}
  {\bibfield  {journal} {\bibinfo  {journal} {\jcap}\ }\textbf {\bibinfo
  {volume} {02}},\ \bibinfo {pages} {056} (\bibinfo {year} {2019})},\ \Eprint
  {https://arxiv.org/abs/1808.07445} {1808.07445} \BibitemShut {NoStop}%
\bibitem [{\citenamefont {{CMB-S4 Collaboration}}(2022)}]{CMBS4:r-forecast}%
  \BibitemOpen
  \bibfield  {author} {\bibinfo {author} {\bibnamefont {{CMB-S4
  Collaboration}}},\ }\href {https://doi.org/10.3847/1538-4357/ac1596}
  {\bibfield  {journal} {\bibinfo  {journal} {\apj}\ }\textbf {\bibinfo
  {volume} {926}},\ \bibinfo {pages} {54} (\bibinfo {year} {2022})},\ \Eprint
  {https://arxiv.org/abs/2008.12619} {2008.12619} \BibitemShut {NoStop}%
\bibitem [{\citenamefont {Sherwin}\ and\ \citenamefont
  {Namikawa}(2023)}]{Sherwin:2021}%
  \BibitemOpen
  \bibfield  {author} {\bibinfo {author} {\bibfnamefont {B.~D.}\ \bibnamefont
  {Sherwin}}\ and\ \bibinfo {author} {\bibfnamefont {T.}~\bibnamefont
  {Namikawa}},\ }\href {https://doi.org/10.1093/mnras/stac3146} {\bibfield
  {journal} {\bibinfo  {journal} {\mnras}\ }\textbf {\bibinfo {volume} {520}},\
  \bibinfo {pages} {3298} (\bibinfo {year} {2023})},\ \Eprint
  {https://arxiv.org/abs/2108.09287} {arXiv:2108.09287 [astro-ph.CO]}
  \BibitemShut {NoStop}%
\bibitem [{\citenamefont {Nakatsuka}\ \emph {et~al.}(2022)\citenamefont
  {Nakatsuka}, \citenamefont {Namikawa},\ and\ \citenamefont
  {Komatsu}}]{Nakatsuka:2022epj}%
  \BibitemOpen
  \bibfield  {author} {\bibinfo {author} {\bibfnamefont {H.}~\bibnamefont
  {Nakatsuka}}, \bibinfo {author} {\bibfnamefont {T.}~\bibnamefont
  {Namikawa}},\ and\ \bibinfo {author} {\bibfnamefont {E.}~\bibnamefont
  {Komatsu}},\ }\href {https://doi.org/10.1103/PhysRevD.105.123509} {\bibfield
  {journal} {\bibinfo  {journal} {\prd}\ }\textbf {\bibinfo {volume} {105}},\
  \bibinfo {pages} {123509} (\bibinfo {year} {2022})},\ \Eprint
  {https://arxiv.org/abs/2203.08560} {arXiv:2203.08560 [astro-ph.CO]}
  \BibitemShut {NoStop}%
\bibitem [{\citenamefont {Galaverni}\ \emph {et~al.}(2023)\citenamefont
  {Galaverni}, \citenamefont {Finelli},\ and\ \citenamefont
  {Paoletti}}]{Galaverni:2023}%
  \BibitemOpen
  \bibfield  {author} {\bibinfo {author} {\bibfnamefont {M.}~\bibnamefont
  {Galaverni}}, \bibinfo {author} {\bibfnamefont {F.}~\bibnamefont {Finelli}},\
  and\ \bibinfo {author} {\bibfnamefont {D.}~\bibnamefont {Paoletti}},\ }\href
  {https://doi.org/10.1103/PhysRevD.107.083529} {\bibfield  {journal} {\bibinfo
   {journal} {\prd}\ }\textbf {\bibinfo {volume} {107}},\ \bibinfo {pages}
  {083529} (\bibinfo {year} {2023})},\ \Eprint
  {https://arxiv.org/abs/2301.07971} {arXiv:2301.07971 [astro-ph.CO]}
  \BibitemShut {NoStop}%
\bibitem [{\citenamefont {Lewis}\ and\ \citenamefont
  {Challinor}(2006)}]{Lewis:2006fu}%
  \BibitemOpen
  \bibfield  {author} {\bibinfo {author} {\bibfnamefont {A.}~\bibnamefont
  {Lewis}}\ and\ \bibinfo {author} {\bibfnamefont {A.}~\bibnamefont
  {Challinor}},\ }\href {https://doi.org/10.1016/j.physrep.2006.03.002}
  {\bibfield  {journal} {\bibinfo  {journal} {Phys. Rept.}\ }\textbf {\bibinfo
  {volume} {429}},\ \bibinfo {pages} {1} (\bibinfo {year} {2006})},\ \Eprint
  {https://arxiv.org/abs/astro-ph/0601594} {arXiv:astro-ph/0601594}
  \BibitemShut {NoStop}%
\bibitem [{\citenamefont {Hanson}\ \emph {et~al.}(2010)\citenamefont {Hanson},
  \citenamefont {Challinor},\ and\ \citenamefont {Lewis}}]{Hanson:2009kr}%
  \BibitemOpen
  \bibfield  {author} {\bibinfo {author} {\bibfnamefont {D.}~\bibnamefont
  {Hanson}}, \bibinfo {author} {\bibfnamefont {A.}~\bibnamefont {Challinor}},\
  and\ \bibinfo {author} {\bibfnamefont {A.}~\bibnamefont {Lewis}},\ }\href
  {https://doi.org/10.1007/s10714-010-1036-y} {\bibfield  {journal} {\bibinfo
  {journal} {Gen. Rel. Grav.}\ }\textbf {\bibinfo {volume} {42}},\ \bibinfo
  {pages} {2197} (\bibinfo {year} {2010})},\ \Eprint
  {https://arxiv.org/abs/0911.0612} {arXiv:0911.0612 [astro-ph.CO]}
  \BibitemShut {NoStop}%
\bibitem [{\citenamefont {Namikawa}(2014)}]{Namikawa:2014xga}%
  \BibitemOpen
  \bibfield  {author} {\bibinfo {author} {\bibfnamefont {T.}~\bibnamefont
  {Namikawa}},\ }\href {https://doi.org/10.1093/ptep/ptu044} {\bibfield
  {journal} {\bibinfo  {journal} {\ptep}\ }\textbf {\bibinfo {volume} {2014}},\
  \bibinfo {pages} {06B108} (\bibinfo {year} {2014})},\ \Eprint
  {https://arxiv.org/abs/1403.3569} {arXiv:1403.3569 [astro-ph.CO]}
  \BibitemShut {NoStop}%
\bibitem [{\citenamefont {Keisler}\ \emph {et~al.}(2011)\citenamefont {Keisler}
  \emph {et~al.}}]{Keisler:2011aw}%
  \BibitemOpen
  \bibfield  {author} {\bibinfo {author} {\bibfnamefont {R.}~\bibnamefont
  {Keisler}} \emph {et~al.},\ }\href
  {https://doi.org/10.1088/0004-637X/743/1/28} {\bibfield  {journal} {\bibinfo
  {journal} {\apj}\ }\textbf {\bibinfo {volume} {743}},\ \bibinfo {pages} {28}
  (\bibinfo {year} {2011})},\ \Eprint {https://arxiv.org/abs/1105.3182}
  {arXiv:1105.3182 [astro-ph.CO]} \BibitemShut {NoStop}%
\bibitem [{\citenamefont {{Choi}}\ \emph {et~al.}(2020)\citenamefont {{Choi}},
  \citenamefont {{Hasselfield}}, \citenamefont {{Ho}}, \citenamefont
  {{Koopman}}, \citenamefont {{Lungu}} \emph {et~al.}}]{ACT:Choi:2020}%
  \BibitemOpen
  \bibfield  {author} {\bibinfo {author} {\bibfnamefont {S.~K.}\ \bibnamefont
  {{Choi}}}, \bibinfo {author} {\bibfnamefont {M.}~\bibnamefont
  {{Hasselfield}}}, \bibinfo {author} {\bibfnamefont {S.-P.~P.}\ \bibnamefont
  {{Ho}}}, \bibinfo {author} {\bibfnamefont {B.}~\bibnamefont {{Koopman}}},
  \bibinfo {author} {\bibfnamefont {M.}~\bibnamefont {{Lungu}}}, \emph
  {et~al.},\ }\href {https://doi.org/10.1088/1475-7516/2020/12/045} {\bibfield
  {journal} {\bibinfo  {journal} {\jcap}\ }\textbf {\bibinfo {volume} {2020}},\
  \bibinfo {pages} {045} (\bibinfo {year} {2020})},\ \Eprint
  {https://arxiv.org/abs/2007.07289} {2007.07289} \BibitemShut {NoStop}%
\bibitem [{\citenamefont {Dutcher}\ \emph {et~al.}(2021)\citenamefont {Dutcher}
  \emph {et~al.}}]{SPT-3G:2021}%
  \BibitemOpen
  \bibfield  {author} {\bibinfo {author} {\bibfnamefont {D.}~\bibnamefont
  {Dutcher}} \emph {et~al.} (\bibinfo {collaboration} {SPT-3G}),\ }\href
  {https://doi.org/10.1103/PhysRevD.104.022003} {\bibfield  {journal} {\bibinfo
   {journal} {\prd}\ }\textbf {\bibinfo {volume} {104}},\ \bibinfo {pages}
  {022003} (\bibinfo {year} {2021})},\ \Eprint
  {https://arxiv.org/abs/2101.01684} {arXiv:2101.01684 [astro-ph.CO]}
  \BibitemShut {NoStop}%
\bibitem [{\citenamefont {Rosenberg}\ \emph {et~al.}(2022)\citenamefont
  {Rosenberg}, \citenamefont {Gratton},\ and\ \citenamefont
  {Efstathiou}}]{Rosenberg:2022:PR4}%
  \BibitemOpen
  \bibfield  {author} {\bibinfo {author} {\bibfnamefont {E.}~\bibnamefont
  {Rosenberg}}, \bibinfo {author} {\bibfnamefont {S.}~\bibnamefont {Gratton}},\
  and\ \bibinfo {author} {\bibfnamefont {G.}~\bibnamefont {Efstathiou}},\
  }\href {https://doi.org/10.1093/mnras/stac2744} {\bibfield  {journal}
  {\bibinfo  {journal} {\mnras}\ }\textbf {\bibinfo {volume} {517}},\ \bibinfo
  {pages} {4620} (\bibinfo {year} {2022})},\ \Eprint
  {https://arxiv.org/abs/2205.10869} {arXiv:2205.10869 [astro-ph.CO]}
  \BibitemShut {NoStop}%
\bibitem [{\citenamefont {Gubitosi}\ \emph {et~al.}(2014)\citenamefont
  {Gubitosi}, \citenamefont {Martinelli},\ and\ \citenamefont
  {Pagano}}]{Gubitosi:2014cua}%
  \BibitemOpen
  \bibfield  {author} {\bibinfo {author} {\bibfnamefont {G.}~\bibnamefont
  {Gubitosi}}, \bibinfo {author} {\bibfnamefont {M.}~\bibnamefont
  {Martinelli}},\ and\ \bibinfo {author} {\bibfnamefont {L.}~\bibnamefont
  {Pagano}},\ }\href {https://doi.org/10.1088/1475-7516/2014/12/020} {\bibfield
   {journal} {\bibinfo  {journal} {\jcap}\ }\textbf {\bibinfo {volume} {12}},\
  \bibinfo {pages} {020} (\bibinfo {year} {2014})},\ \Eprint
  {https://arxiv.org/abs/1410.1799} {arXiv:1410.1799 [astro-ph.CO]}
  \BibitemShut {NoStop}%
\bibitem [{\citenamefont {Challinor}\ and\ \citenamefont
  {Lewis}(2005)}]{Challinor:2005jy}%
  \BibitemOpen
  \bibfield  {author} {\bibinfo {author} {\bibfnamefont {A.}~\bibnamefont
  {Challinor}}\ and\ \bibinfo {author} {\bibfnamefont {A.}~\bibnamefont
  {Lewis}},\ }\href {https://doi.org/10.1103/PhysRevD.71.103010} {\bibfield
  {journal} {\bibinfo  {journal} {\prd}\ }\textbf {\bibinfo {volume} {71}},\
  \bibinfo {pages} {103010} (\bibinfo {year} {2005})},\ \Eprint
  {https://arxiv.org/abs/astro-ph/0502425} {arXiv:astro-ph/0502425}
  \BibitemShut {NoStop}%
\bibitem [{\citenamefont {Zaldarriaga}\ and\ \citenamefont
  {Seljak}(1997)}]{Zaldarriaga:1996xe}%
  \BibitemOpen
  \bibfield  {author} {\bibinfo {author} {\bibfnamefont {M.}~\bibnamefont
  {Zaldarriaga}}\ and\ \bibinfo {author} {\bibfnamefont {U.}~\bibnamefont
  {Seljak}},\ }\href {https://doi.org/10.1103/PhysRevD.55.1830} {\bibfield
  {journal} {\bibinfo  {journal} {\prd}\ }\textbf {\bibinfo {volume} {55}},\
  \bibinfo {pages} {1830} (\bibinfo {year} {1997})},\ \Eprint
  {https://arxiv.org/abs/astro-ph/9609170} {arXiv:astro-ph/9609170}
  \BibitemShut {NoStop}%
\bibitem [{\citenamefont {Kamionkowski}\ \emph {et~al.}(1997)\citenamefont
  {Kamionkowski}, \citenamefont {Kosowsky},\ and\ \citenamefont
  {Stebbins}}]{Kamionkowski:1996ks}%
  \BibitemOpen
  \bibfield  {author} {\bibinfo {author} {\bibfnamefont {M.}~\bibnamefont
  {Kamionkowski}}, \bibinfo {author} {\bibfnamefont {A.}~\bibnamefont
  {Kosowsky}},\ and\ \bibinfo {author} {\bibfnamefont {A.}~\bibnamefont
  {Stebbins}},\ }\href {https://doi.org/10.1103/PhysRevD.55.7368} {\bibfield
  {journal} {\bibinfo  {journal} {Phys. Rev. D}\ }\textbf {\bibinfo {volume}
  {55}},\ \bibinfo {pages} {7368} (\bibinfo {year} {1997})},\ \Eprint
  {https://arxiv.org/abs/astro-ph/9611125} {arXiv:astro-ph/9611125}
  \BibitemShut {NoStop}%
\bibitem [{\citenamefont {Zhao}\ \emph {et~al.}(2015)\citenamefont {Zhao},
  \citenamefont {Wang}, \citenamefont {Xia}, \citenamefont {Li},\ and\
  \citenamefont {Zhang}}]{Zhao:2015mqa}%
  \BibitemOpen
  \bibfield  {author} {\bibinfo {author} {\bibfnamefont {G.-B.}\ \bibnamefont
  {Zhao}}, \bibinfo {author} {\bibfnamefont {Y.}~\bibnamefont {Wang}}, \bibinfo
  {author} {\bibfnamefont {J.-Q.}\ \bibnamefont {Xia}}, \bibinfo {author}
  {\bibfnamefont {M.}~\bibnamefont {Li}},\ and\ \bibinfo {author}
  {\bibfnamefont {X.}~\bibnamefont {Zhang}},\ }\href
  {https://doi.org/10.1088/1475-7516/2015/07/032} {\bibfield  {journal}
  {\bibinfo  {journal} {\jcap}\ }\textbf {\bibinfo {volume} {07}},\ \bibinfo
  {pages} {032} (\bibinfo {year} {2015})},\ \Eprint
  {https://arxiv.org/abs/1504.04507} {arXiv:1504.04507 [astro-ph.CO]}
  \BibitemShut {NoStop}%
\bibitem [{\citenamefont {Wu}\ \emph {et~al.}(2009)\citenamefont {Wu} \emph
  {et~al.}}]{QUaD:2008ado}%
  \BibitemOpen
  \bibfield  {author} {\bibinfo {author} {\bibfnamefont {E.~Y.~S.}\
  \bibnamefont {Wu}} \emph {et~al.} (\bibinfo {collaboration} {QUaD}),\ }\href
  {https://doi.org/10.1103/PhysRevLett.102.161302} {\bibfield  {journal}
  {\bibinfo  {journal} {Phys. Rev. Lett.}\ }\textbf {\bibinfo {volume} {102}},\
  \bibinfo {pages} {161302} (\bibinfo {year} {2009})},\ \Eprint
  {https://arxiv.org/abs/0811.0618} {arXiv:0811.0618 [astro-ph]} \BibitemShut
  {NoStop}%
\bibitem [{\citenamefont {Miller}\ \emph {et~al.}(2009)\citenamefont {Miller},
  \citenamefont {Shimon},\ and\ \citenamefont {Keating}}]{Miller:2009pt}%
  \BibitemOpen
  \bibfield  {author} {\bibinfo {author} {\bibfnamefont {N.~J.}\ \bibnamefont
  {Miller}}, \bibinfo {author} {\bibfnamefont {M.}~\bibnamefont {Shimon}},\
  and\ \bibinfo {author} {\bibfnamefont {B.~G.}\ \bibnamefont {Keating}},\
  }\href {https://doi.org/10.1103/PhysRevD.79.103002} {\bibfield  {journal}
  {\bibinfo  {journal} {Phys. Rev. D}\ }\textbf {\bibinfo {volume} {79}},\
  \bibinfo {pages} {103002} (\bibinfo {year} {2009})},\ \Eprint
  {https://arxiv.org/abs/0903.1116} {arXiv:0903.1116 [astro-ph.CO]}
  \BibitemShut {NoStop}%
\bibitem [{\citenamefont {Komatsu}\ \emph {et~al.}(2011)\citenamefont {Komatsu}
  \emph {et~al.}}]{Komatsu:2010:WMAP7}%
  \BibitemOpen
  \bibfield  {author} {\bibinfo {author} {\bibfnamefont {E.}~\bibnamefont
  {Komatsu}} \emph {et~al.},\ }\href
  {https://doi.org/10.1088/0067-0049/192/2/18} {\bibfield  {journal} {\bibinfo
  {journal} {\apj}\ }\textbf {\bibinfo {volume} {192}},\ \bibinfo {pages} {18}
  (\bibinfo {year} {2011})},\ \Eprint {https://arxiv.org/abs/1001.4538}
  {1001.4538} \BibitemShut {NoStop}%
\bibitem [{\citenamefont {{Keating}}\ \emph {et~al.}(2013)\citenamefont
  {{Keating}}, \citenamefont {{Shimon}},\ and\ \citenamefont
  {{Yadav}}}]{Keating:2013}%
  \BibitemOpen
  \bibfield  {author} {\bibinfo {author} {\bibfnamefont {B.~G.}\ \bibnamefont
  {{Keating}}}, \bibinfo {author} {\bibfnamefont {M.}~\bibnamefont
  {{Shimon}}},\ and\ \bibinfo {author} {\bibfnamefont {A.~P.~S.}\ \bibnamefont
  {{Yadav}}},\ }\href {https://doi.org/10.1088/2041-8205/762/2/L23} {\bibfield
  {journal} {\bibinfo  {journal} {\apjl}\ }\textbf {\bibinfo {volume} {762}},\
  \bibinfo {pages} {L23} (\bibinfo {year} {2013})},\ \Eprint
  {https://arxiv.org/abs/1211.5734} {1211.5734} \BibitemShut {NoStop}%
\bibitem [{\citenamefont {Minami}\ \emph {et~al.}(2019)\citenamefont {Minami},
  \citenamefont {Ochi}, \citenamefont {Ichiki}, \citenamefont {Katayama},
  \citenamefont {Komatsu},\ and\ \citenamefont {Matsumura}}]{Minami:2019ruj}%
  \BibitemOpen
  \bibfield  {author} {\bibinfo {author} {\bibfnamefont {Y.}~\bibnamefont
  {Minami}}, \bibinfo {author} {\bibfnamefont {H.}~\bibnamefont {Ochi}},
  \bibinfo {author} {\bibfnamefont {K.}~\bibnamefont {Ichiki}}, \bibinfo
  {author} {\bibfnamefont {N.}~\bibnamefont {Katayama}}, \bibinfo {author}
  {\bibfnamefont {E.}~\bibnamefont {Komatsu}},\ and\ \bibinfo {author}
  {\bibfnamefont {T.}~\bibnamefont {Matsumura}},\ }\href
  {https://doi.org/10.1093/ptep/ptz079} {\bibfield  {journal} {\bibinfo
  {journal} {\ptep}\ }\textbf {\bibinfo {volume} {2019}},\ \bibinfo {pages}
  {083E02} (\bibinfo {year} {2019})},\ \Eprint
  {https://arxiv.org/abs/1904.12440} {arXiv:1904.12440 [astro-ph.CO]}
  \BibitemShut {NoStop}%
\bibitem [{\citenamefont {Aumont}\ \emph {et~al.}(2020)\citenamefont {Aumont},
  \citenamefont {Mac\'\i{}as-P\'erez}, \citenamefont {Ritacco}, \citenamefont
  {Ponthieu},\ and\ \citenamefont {Mangilli}}]{Aumont:2018epb}%
  \BibitemOpen
  \bibfield  {author} {\bibinfo {author} {\bibfnamefont {J.}~\bibnamefont
  {Aumont}}, \bibinfo {author} {\bibfnamefont {J.~F.}\ \bibnamefont
  {Mac\'\i{}as-P\'erez}}, \bibinfo {author} {\bibfnamefont {A.}~\bibnamefont
  {Ritacco}}, \bibinfo {author} {\bibfnamefont {N.}~\bibnamefont {Ponthieu}},\
  and\ \bibinfo {author} {\bibfnamefont {A.}~\bibnamefont {Mangilli}},\ }\href
  {https://doi.org/10.1051/0004-6361/201833504} {\bibfield  {journal} {\bibinfo
   {journal} {Astron. Astrophys.}\ }\textbf {\bibinfo {volume} {634}},\
  \bibinfo {pages} {A100} (\bibinfo {year} {2020})},\ \Eprint
  {https://arxiv.org/abs/1805.10475} {arXiv:1805.10475 [astro-ph.CO]}
  \BibitemShut {NoStop}%
\bibitem [{\citenamefont {Cornelison}\ \emph {et~al.}(2022)\citenamefont
  {Cornelison} \emph {et~al.}}]{Cornelison:2022zrc}%
  \BibitemOpen
  \bibfield  {author} {\bibinfo {author} {\bibfnamefont {J.}~\bibnamefont
  {Cornelison}} \emph {et~al.},\ }\href {https://doi.org/10.1117/12.2620212}
  {\bibfield  {journal} {\bibinfo  {journal} {Proc. SPIE Int. Soc. Opt. Eng.}\
  }\textbf {\bibinfo {volume} {12190}},\ \bibinfo {pages} {121901X} (\bibinfo
  {year} {2022})},\ \Eprint {https://arxiv.org/abs/2207.14796}
  {arXiv:2207.14796 [astro-ph.IM]} \BibitemShut {NoStop}%
\bibitem [{\citenamefont {Liu}\ \emph {et~al.}(2006)\citenamefont {Liu},
  \citenamefont {Lee},\ and\ \citenamefont {Ng}}]{Liu:2006uh}%
  \BibitemOpen
  \bibfield  {author} {\bibinfo {author} {\bibfnamefont {G.-C.}\ \bibnamefont
  {Liu}}, \bibinfo {author} {\bibfnamefont {S.}~\bibnamefont {Lee}},\ and\
  \bibinfo {author} {\bibfnamefont {K.-W.}\ \bibnamefont {Ng}},\ }\href
  {https://doi.org/10.1103/PhysRevLett.97.161303} {\bibfield  {journal}
  {\bibinfo  {journal} {\prl}\ }\textbf {\bibinfo {volume} {97}},\ \bibinfo
  {pages} {161303} (\bibinfo {year} {2006})},\ \Eprint
  {https://arxiv.org/abs/astro-ph/0606248} {arXiv:astro-ph/0606248}
  \BibitemShut {NoStop}%
\bibitem [{\citenamefont {Lee}\ \emph {et~al.}(2016)\citenamefont {Lee},
  \citenamefont {Liu},\ and\ \citenamefont {Ng}}]{Lee:2016jym}%
  \BibitemOpen
  \bibfield  {author} {\bibinfo {author} {\bibfnamefont {S.}~\bibnamefont
  {Lee}}, \bibinfo {author} {\bibfnamefont {G.-C.}\ \bibnamefont {Liu}},\ and\
  \bibinfo {author} {\bibfnamefont {K.-W.}\ \bibnamefont {Ng}},\ }\href@noop {}
  {\bibfield  {journal} {\bibinfo  {journal} {The Universe}\ }\textbf {\bibinfo
  {volume} {4}},\ \bibinfo {pages} {29} (\bibinfo {year} {2016})},\ \Eprint
  {https://arxiv.org/abs/1912.12903} {arXiv:1912.12903 [astro-ph.CO]}
  \BibitemShut {NoStop}%
\bibitem [{\citenamefont {{Lesgourgues}}(2011)}]{CLASS}%
  \BibitemOpen
  \bibfield  {author} {\bibinfo {author} {\bibfnamefont {J.}~\bibnamefont
  {{Lesgourgues}}},\ }\href {https://doi.org/10.48550/arXiv.1104.2932}
  {\bibfield  {journal} {\bibinfo  {journal} {arXiv e-prints}\ ,\ \bibinfo
  {eid} {arXiv:1104.2932}} (\bibinfo {year} {2011})},\ \Eprint
  {https://arxiv.org/abs/1104.2932} {1104.2932} \BibitemShut {NoStop}%
\bibitem [{\citenamefont {Seljak}(1996)}]{Seljak:1995}%
  \BibitemOpen
  \bibfield  {author} {\bibinfo {author} {\bibfnamefont {U.}~\bibnamefont
  {Seljak}},\ }\href {https://doi.org/10.1086/177218} {\bibfield  {journal}
  {\bibinfo  {journal} {\apj}\ }\textbf {\bibinfo {volume} {463}},\ \bibinfo
  {pages} {1} (\bibinfo {year} {1996})},\ \Eprint
  {https://arxiv.org/abs/astro-ph/9505109} {arXiv:astro-ph/9505109}
  \BibitemShut {NoStop}%
\bibitem [{\citenamefont {Chon}\ \emph {et~al.}(2004)\citenamefont {Chon},
  \citenamefont {Challinor}, \citenamefont {Prunet}, \citenamefont {Hivon},\
  and\ \citenamefont {Szapudi}}]{Chon:2003gx}%
  \BibitemOpen
  \bibfield  {author} {\bibinfo {author} {\bibfnamefont {G.}~\bibnamefont
  {Chon}}, \bibinfo {author} {\bibfnamefont {A.}~\bibnamefont {Challinor}},
  \bibinfo {author} {\bibfnamefont {S.}~\bibnamefont {Prunet}}, \bibinfo
  {author} {\bibfnamefont {E.}~\bibnamefont {Hivon}},\ and\ \bibinfo {author}
  {\bibfnamefont {I.}~\bibnamefont {Szapudi}},\ }\href
  {https://doi.org/10.1111/j.1365-2966.2004.07737.x} {\bibfield  {journal}
  {\bibinfo  {journal} {\mnras}\ }\textbf {\bibinfo {volume} {350}},\ \bibinfo
  {pages} {914} (\bibinfo {year} {2004})},\ \Eprint
  {https://arxiv.org/abs/astro-ph/0303414} {arXiv:astro-ph/0303414}
  \BibitemShut {NoStop}%
\bibitem [{\citenamefont {Namikawa}(2021)}]{Namikawa:2021:mode}%
  \BibitemOpen
  \bibfield  {author} {\bibinfo {author} {\bibfnamefont {T.}~\bibnamefont
  {Namikawa}},\ }\href {https://doi.org/10.1093/mnras/stab1796} {\bibfield
  {journal} {\bibinfo  {journal} {\mnras}\ }\textbf {\bibinfo {volume} {506}},\
  \bibinfo {pages} {1250} (\bibinfo {year} {2021})},\ \Eprint
  {https://arxiv.org/abs/2105.03367} {arXiv:2105.03367 [astro-ph.CO]}
  \BibitemShut {NoStop}%
\bibitem [{\citenamefont {Saito}\ \emph {et~al.}(2007)\citenamefont {Saito},
  \citenamefont {Ichiki},\ and\ \citenamefont {Taruya}}]{Saito:2007kt}%
  \BibitemOpen
  \bibfield  {author} {\bibinfo {author} {\bibfnamefont {S.}~\bibnamefont
  {Saito}}, \bibinfo {author} {\bibfnamefont {K.}~\bibnamefont {Ichiki}},\ and\
  \bibinfo {author} {\bibfnamefont {A.}~\bibnamefont {Taruya}},\ }\href
  {https://doi.org/10.1088/1475-7516/2007/09/002} {\bibfield  {journal}
  {\bibinfo  {journal} {\jcap}\ }\textbf {\bibinfo {volume} {09}},\ \bibinfo
  {pages} {002} (\bibinfo {year} {2007})},\ \Eprint
  {https://arxiv.org/abs/0705.3701} {arXiv:0705.3701 [astro-ph]} \BibitemShut
  {NoStop}%
\bibitem [{\citenamefont {Katayama}\ and\ \citenamefont
  {Komatsu}(2011)}]{Katayama:2011eh}%
  \BibitemOpen
  \bibfield  {author} {\bibinfo {author} {\bibfnamefont {N.}~\bibnamefont
  {Katayama}}\ and\ \bibinfo {author} {\bibfnamefont {E.}~\bibnamefont
  {Komatsu}},\ }\href@noop {} {\bibfield  {journal} {\bibinfo  {journal}
  {\apj}\ }\textbf {\bibinfo {volume} {737}},\ \bibinfo {pages} {78} (\bibinfo
  {year} {2011})},\ \Eprint {https://arxiv.org/abs/1101.5210} {1101.5210}
  \BibitemShut {NoStop}%
\bibitem [{\citenamefont {Gupta}\ and\ \citenamefont
  {Nagar}(1999)}]{Gupta&Nagar:1999}%
  \BibitemOpen
  \bibfield  {author} {\bibinfo {author} {\bibfnamefont {A.}~\bibnamefont
  {Gupta}}\ and\ \bibinfo {author} {\bibfnamefont {D.}~\bibnamefont {Nagar}},\
  }\href {https://doi.org/10.1201/9780203749289} {\emph {\bibinfo {title}
  {Matrix Variate Distributions}}}\ (\bibinfo  {publisher} {Chapman and
  Hall/CRC},\ \bibinfo {year} {1999})\BibitemShut {NoStop}%
\bibitem [{\citenamefont {Hamimeche}\ and\ \citenamefont
  {Lewis}(2008)}]{Hamimeche:2008ai}%
  \BibitemOpen
  \bibfield  {author} {\bibinfo {author} {\bibfnamefont {S.}~\bibnamefont
  {Hamimeche}}\ and\ \bibinfo {author} {\bibfnamefont {A.}~\bibnamefont
  {Lewis}},\ }\href {https://doi.org/10.1103/PhysRevD.77.103013} {\bibfield
  {journal} {\bibinfo  {journal} {\prd}\ }\textbf {\bibinfo {volume} {77}},\
  \bibinfo {pages} {103013} (\bibinfo {year} {2008})},\ \Eprint
  {https://arxiv.org/abs/0801.0554} {arXiv:0801.0554 [astro-ph]} \BibitemShut
  {NoStop}%
\bibitem [{\citenamefont {Johnson}\ \emph {et~al.}(2015)\citenamefont
  {Johnson}, \citenamefont {Vourch}, \citenamefont {Drysdale}, \citenamefont
  {Kalman}, \citenamefont {Fujikawa}, \citenamefont {Keating},\ and\
  \citenamefont {Kaufman}}]{Johnson_2015}%
  \BibitemOpen
  \bibfield  {author} {\bibinfo {author} {\bibfnamefont {B.~R.}\ \bibnamefont
  {Johnson}}, \bibinfo {author} {\bibfnamefont {C.~J.}\ \bibnamefont {Vourch}},
  \bibinfo {author} {\bibfnamefont {T.~D.}\ \bibnamefont {Drysdale}}, \bibinfo
  {author} {\bibfnamefont {A.}~\bibnamefont {Kalman}}, \bibinfo {author}
  {\bibfnamefont {S.}~\bibnamefont {Fujikawa}}, \bibinfo {author}
  {\bibfnamefont {B.}~\bibnamefont {Keating}},\ and\ \bibinfo {author}
  {\bibfnamefont {J.}~\bibnamefont {Kaufman}},\ }\bibfield  {journal} {\bibinfo
   {journal} {J. Astron. Inst.}\ }\textbf {\bibinfo {volume} {04}},\ \href
  {https://doi.org/10.1142/s2251171715500075} {10.1142/s2251171715500075}
  (\bibinfo {year} {2015})\BibitemShut {NoStop}%
\bibitem [{\citenamefont {Nati}\ \emph {et~al.}(2017)\citenamefont {Nati},
  \citenamefont {Devlin}, \citenamefont {Gerbino}, \citenamefont {Johnson},
  \citenamefont {Keating}, \citenamefont {Pagano},\ and\ \citenamefont
  {Teply}}]{Nati:2017lnn}%
  \BibitemOpen
  \bibfield  {author} {\bibinfo {author} {\bibfnamefont {F.}~\bibnamefont
  {Nati}}, \bibinfo {author} {\bibfnamefont {M.~J.}\ \bibnamefont {Devlin}},
  \bibinfo {author} {\bibfnamefont {M.}~\bibnamefont {Gerbino}}, \bibinfo
  {author} {\bibfnamefont {B.~R.}\ \bibnamefont {Johnson}}, \bibinfo {author}
  {\bibfnamefont {B.}~\bibnamefont {Keating}}, \bibinfo {author} {\bibfnamefont
  {L.}~\bibnamefont {Pagano}},\ and\ \bibinfo {author} {\bibfnamefont
  {G.}~\bibnamefont {Teply}},\ }\href
  {https://doi.org/10.1142/S2251171717400086} {\bibfield  {journal} {\bibinfo
  {journal} {J. Astron. Inst.}\ }\textbf {\bibinfo {volume} {06}},\ \bibinfo
  {pages} {1740008} (\bibinfo {year} {2017})},\ \Eprint
  {https://arxiv.org/abs/1704.02704} {arXiv:1704.02704 [astro-ph.IM]}
  \BibitemShut {NoStop}%
\bibitem [{\citenamefont {Casas}\ \emph {et~al.}(2021)\citenamefont {Casas},
  \citenamefont {Martínez-González}, \citenamefont {Bermejo-Ballesteros},
  \citenamefont {García}, \citenamefont {Cubas}, \citenamefont {Vielva},
  \citenamefont {Barreiro},\ and\ \citenamefont {Sanz}}]{Casas}%
  \BibitemOpen
  \bibfield  {author} {\bibinfo {author} {\bibfnamefont {F.~J.}\ \bibnamefont
  {Casas}}, \bibinfo {author} {\bibfnamefont {E.}~\bibnamefont
  {Martínez-González}}, \bibinfo {author} {\bibfnamefont {J.}~\bibnamefont
  {Bermejo-Ballesteros}}, \bibinfo {author} {\bibfnamefont {S.}~\bibnamefont
  {García}}, \bibinfo {author} {\bibfnamefont {J.}~\bibnamefont {Cubas}},
  \bibinfo {author} {\bibfnamefont {P.}~\bibnamefont {Vielva}}, \bibinfo
  {author} {\bibfnamefont {R.~B.}\ \bibnamefont {Barreiro}},\ and\ \bibinfo
  {author} {\bibfnamefont {A.}~\bibnamefont {Sanz}},\ }\bibfield  {journal}
  {\bibinfo  {journal} {Sensors}\ }\textbf {\bibinfo {volume} {21}},\ \href
  {https://doi.org/10.3390/s21103361} {10.3390/s21103361} (\bibinfo {year}
  {2021})\BibitemShut {NoStop}%
\bibitem [{\citenamefont {Mirmelstein}\ \emph {et~al.}(2021)\citenamefont
  {Mirmelstein}, \citenamefont {Fabbian}, \citenamefont {Lewis},\ and\
  \citenamefont {Peloton}}]{Mirmelstein:2020}%
  \BibitemOpen
  \bibfield  {author} {\bibinfo {author} {\bibfnamefont {M.}~\bibnamefont
  {Mirmelstein}}, \bibinfo {author} {\bibfnamefont {G.}~\bibnamefont
  {Fabbian}}, \bibinfo {author} {\bibfnamefont {A.}~\bibnamefont {Lewis}},\
  and\ \bibinfo {author} {\bibfnamefont {J.}~\bibnamefont {Peloton}},\ }\href
  {https://doi.org/10.1103/PhysRevD.103.123540} {\bibfield  {journal} {\bibinfo
   {journal} {\prd}\ }\textbf {\bibinfo {volume} {103}},\ \bibinfo {pages}
  {123540} (\bibinfo {year} {2021})},\ \Eprint
  {https://arxiv.org/abs/2011.13910} {arXiv:2011.13910 [astro-ph.CO]}
  \BibitemShut {NoStop}%
\bibitem [{\citenamefont {Nagata}\ and\ \citenamefont
  {Namikawa}(2021)}]{Nagata:2021}%
  \BibitemOpen
  \bibfield  {author} {\bibinfo {author} {\bibfnamefont {R.}~\bibnamefont
  {Nagata}}\ and\ \bibinfo {author} {\bibfnamefont {T.}~\bibnamefont
  {Namikawa}},\ }\href {https://doi.org/10.1093/ptep/ptab040} {\bibfield
  {journal} {\bibinfo  {journal} {\ptep}\ }\textbf {\bibinfo {volume} {2021}},\
  \bibinfo {pages} {053} (\bibinfo {year} {2021})},\ \Eprint
  {https://arxiv.org/abs/2102.00133} {arXiv:2102.00133 [astro-ph.CO]}
  \BibitemShut {NoStop}%
\bibitem [{\citenamefont {Thorne}\ \emph {et~al.}(2018)\citenamefont {Thorne},
  \citenamefont {Fujita}, \citenamefont {Hazumi}, \citenamefont {Katayama},
  \citenamefont {Komatsu},\ and\ \citenamefont {Shiraishi}}]{Thorne:2017jft}%
  \BibitemOpen
  \bibfield  {author} {\bibinfo {author} {\bibfnamefont {B.}~\bibnamefont
  {Thorne}}, \bibinfo {author} {\bibfnamefont {T.}~\bibnamefont {Fujita}},
  \bibinfo {author} {\bibfnamefont {M.}~\bibnamefont {Hazumi}}, \bibinfo
  {author} {\bibfnamefont {N.}~\bibnamefont {Katayama}}, \bibinfo {author}
  {\bibfnamefont {E.}~\bibnamefont {Komatsu}},\ and\ \bibinfo {author}
  {\bibfnamefont {M.}~\bibnamefont {Shiraishi}},\ }\href
  {https://doi.org/10.1103/PhysRevD.97.043506} {\bibfield  {journal} {\bibinfo
  {journal} {\prd}\ }\textbf {\bibinfo {volume} {97}},\ \bibinfo {pages}
  {043506} (\bibinfo {year} {2018})},\ \Eprint
  {https://arxiv.org/abs/1707.03240} {arXiv:1707.03240 [astro-ph.CO]}
  \BibitemShut {NoStop}%
\bibitem [{\citenamefont {Fujita}\ \emph {et~al.}(2022)\citenamefont {Fujita},
  \citenamefont {Minami}, \citenamefont {Shiraishi},\ and\ \citenamefont
  {Yokoyama}}]{Fujita:2022qlk}%
  \BibitemOpen
  \bibfield  {author} {\bibinfo {author} {\bibfnamefont {T.}~\bibnamefont
  {Fujita}}, \bibinfo {author} {\bibfnamefont {Y.}~\bibnamefont {Minami}},
  \bibinfo {author} {\bibfnamefont {M.}~\bibnamefont {Shiraishi}},\ and\
  \bibinfo {author} {\bibfnamefont {S.}~\bibnamefont {Yokoyama}},\ }\href
  {https://doi.org/10.1103/PhysRevD.106.103529} {\bibfield  {journal} {\bibinfo
   {journal} {\prd}\ }\textbf {\bibinfo {volume} {106}},\ \bibinfo {pages}
  {103529} (\bibinfo {year} {2022})},\ \Eprint
  {https://arxiv.org/abs/2208.08101} {arXiv:2208.08101 [astro-ph.CO]}
  \BibitemShut {NoStop}%
\bibitem [{\citenamefont {Gonzalez}\ \emph {et~al.}(2022)\citenamefont
  {Gonzalez}, \citenamefont {Kitajima}, \citenamefont {Takahashi},\ and\
  \citenamefont {Yin}}]{Gonzalez:2022mcx}%
  \BibitemOpen
  \bibfield  {author} {\bibinfo {author} {\bibfnamefont {D.}~\bibnamefont
  {Gonzalez}}, \bibinfo {author} {\bibfnamefont {N.}~\bibnamefont {Kitajima}},
  \bibinfo {author} {\bibfnamefont {F.}~\bibnamefont {Takahashi}},\ and\
  \bibinfo {author} {\bibfnamefont {W.}~\bibnamefont {Yin}},\ }\href@noop {}
  {\bibfield  {journal} {\bibinfo  {journal} {{ }}\ } (\bibinfo {year}
  {2022})},\ \Eprint {https://arxiv.org/abs/2211.06849} {arXiv:2211.06849
  [hep-ph]} \BibitemShut {NoStop}%
\bibitem [{\citenamefont {Gasparotto}\ and\ \citenamefont
  {Obata}(2022)}]{Gasparotto:2022uqo}%
  \BibitemOpen
  \bibfield  {author} {\bibinfo {author} {\bibfnamefont {S.}~\bibnamefont
  {Gasparotto}}\ and\ \bibinfo {author} {\bibfnamefont {I.}~\bibnamefont
  {Obata}},\ }\href {https://doi.org/10.1088/1475-7516/2022/08/025} {\bibfield
  {journal} {\bibinfo  {journal} {\jcap}\ }\textbf {\bibinfo {volume} {08}},\
  \bibinfo {pages} {025} (\bibinfo {year} {2022})},\ \Eprint
  {https://arxiv.org/abs/2203.09409} {arXiv:2203.09409 [astro-ph.CO]}
  \BibitemShut {NoStop}%
\bibitem [{\citenamefont {Obata}(2022)}]{Obata:2021nql}%
  \BibitemOpen
  \bibfield  {author} {\bibinfo {author} {\bibfnamefont {I.}~\bibnamefont
  {Obata}},\ }\href {https://doi.org/10.1088/1475-7516/2022/09/062} {\bibfield
  {journal} {\bibinfo  {journal} {\jcap}\ }\textbf {\bibinfo {volume} {09}},\
  \bibinfo {pages} {062} (\bibinfo {year} {2022})},\ \Eprint
  {https://arxiv.org/abs/2108.02150} {arXiv:2108.02150 [astro-ph.CO]}
  \BibitemShut {NoStop}%
\bibitem [{\citenamefont {Murai}\ \emph {et~al.}(2023)\citenamefont {Murai},
  \citenamefont {Naokawa}, \citenamefont {Namikawa},\ and\ \citenamefont
  {Komatsu}}]{Murai:2022:EDE}%
  \BibitemOpen
  \bibfield  {author} {\bibinfo {author} {\bibfnamefont {K.}~\bibnamefont
  {Murai}}, \bibinfo {author} {\bibfnamefont {F.}~\bibnamefont {Naokawa}},
  \bibinfo {author} {\bibfnamefont {T.}~\bibnamefont {Namikawa}},\ and\
  \bibinfo {author} {\bibfnamefont {E.}~\bibnamefont {Komatsu}},\ }\href
  {https://doi.org/10.1103/PhysRevD.107.L041302} {\bibfield  {journal}
  {\bibinfo  {journal} {\prd}\ }\textbf {\bibinfo {volume} {107}},\ \bibinfo
  {pages} {L041302} (\bibinfo {year} {2023})},\ \Eprint
  {https://arxiv.org/abs/2209.07804} {arXiv:2209.07804 [astro-ph.CO]}
  \BibitemShut {NoStop}%
\bibitem [{\citenamefont {Eskilt}\ \emph {et~al.}(2023)\citenamefont {Eskilt},
  \citenamefont {Herold}, \citenamefont {Komatsu}, \citenamefont {Murai},
  \citenamefont {Namikawa},\ and\ \citenamefont {Naokawa}}]{Eskilt:2023:EDE}%
  \BibitemOpen
  \bibfield  {author} {\bibinfo {author} {\bibfnamefont {J.~R.}\ \bibnamefont
  {Eskilt}}, \bibinfo {author} {\bibfnamefont {L.}~\bibnamefont {Herold}},
  \bibinfo {author} {\bibfnamefont {E.}~\bibnamefont {Komatsu}}, \bibinfo
  {author} {\bibfnamefont {K.}~\bibnamefont {Murai}}, \bibinfo {author}
  {\bibfnamefont {T.}~\bibnamefont {Namikawa}},\ and\ \bibinfo {author}
  {\bibfnamefont {F.}~\bibnamefont {Naokawa}},\ }\href@noop {} {\bibfield
  {journal} {\bibinfo  {journal} {{}}\ } (\bibinfo {year} {2023})},\ \Eprint
  {https://arxiv.org/abs/2303.15369} {arXiv:2303.15369 [astro-ph.CO]}
  \BibitemShut {NoStop}%
\bibitem [{\citenamefont {Yin}\ \emph {et~al.}(2023)\citenamefont {Yin},
  \citenamefont {Kochappan}, \citenamefont {Ghosh},\ and\ \citenamefont
  {Lee}}]{Yin:2023}%
  \BibitemOpen
  \bibfield  {author} {\bibinfo {author} {\bibfnamefont {L.}~\bibnamefont
  {Yin}}, \bibinfo {author} {\bibfnamefont {J.}~\bibnamefont {Kochappan}},
  \bibinfo {author} {\bibfnamefont {T.}~\bibnamefont {Ghosh}},\ and\ \bibinfo
  {author} {\bibfnamefont {B.-H.}\ \bibnamefont {Lee}},\ }\href@noop {}
  {\bibfield  {journal} {\bibinfo  {journal} {{ }}\ } (\bibinfo {year}
  {2023})},\ \Eprint {https://arxiv.org/abs/2305.07937} {2305.07937}
  \BibitemShut {NoStop}%
\bibitem [{\citenamefont {Moncelsi}\ \emph {et~al.}(2020)\citenamefont
  {Moncelsi} \emph {et~al.}}]{Moncelsi:2020ppj}%
  \BibitemOpen
  \bibfield  {author} {\bibinfo {author} {\bibfnamefont {L.}~\bibnamefont
  {Moncelsi}} \emph {et~al.},\ }\href {https://doi.org/10.1117/12.2561995}
  {\bibfield  {journal} {\bibinfo  {journal} {Proc. SPIE Int. Soc. Opt. Eng.}\
  }\textbf {\bibinfo {volume} {11453}},\ \bibinfo {pages} {1145314} (\bibinfo
  {year} {2020})},\ \Eprint {https://arxiv.org/abs/2012.04047}
  {arXiv:2012.04047 [astro-ph.IM]} \BibitemShut {NoStop}%
\bibitem [{\citenamefont {Suzuki}\ \emph {et~al.}(2016)\citenamefont {Suzuki}
  \emph {et~al.}}]{POLARBEAR:2015ixw}%
  \BibitemOpen
  \bibfield  {author} {\bibinfo {author} {\bibfnamefont {A.}~\bibnamefont
  {Suzuki}} \emph {et~al.} (\bibinfo {collaboration} {POLARBEAR}),\ }\href
  {https://doi.org/10.1007/s10909-015-1425-4} {\bibfield  {journal} {\bibinfo
  {journal} {J. Low Temp. Phys.}\ }\textbf {\bibinfo {volume} {184}},\ \bibinfo
  {pages} {805} (\bibinfo {year} {2016})},\ \Eprint
  {https://arxiv.org/abs/1512.07299} {arXiv:1512.07299 [astro-ph.IM]}
  \BibitemShut {NoStop}%
\bibitem [{\citenamefont {{Li}}\ \emph {et~al.}(2021)\citenamefont {{Li}},
  \citenamefont {{Austermann}}, \citenamefont {{Beall}}, \citenamefont
  {{Bruno}}, \citenamefont {{Choi}}, \citenamefont {{Cothard}}, \citenamefont
  {{Crowley}}, \citenamefont {{Duff}}, \citenamefont {{Ho}}, \citenamefont
  {{Golec}}, \citenamefont {{Hilton}}, \citenamefont {{Hasselfield}},
  \citenamefont {{Hubmayr}}, \citenamefont {{Koopman}}, \citenamefont
  {{Lungu}}, \citenamefont {{McMahon}}, \citenamefont {{Niemack}},
  \citenamefont {{Page}}, \citenamefont {{Salatino}}, \citenamefont {{Simon}},
  \citenamefont {{Staggs}}, \citenamefont {{Stevens}}, \citenamefont {{Ullom}},
  \citenamefont {{Vavagiakis}}, \citenamefont {{Wang}}, \citenamefont
  {{Wollack}},\ and\ \citenamefont {{Xu}}}]{2021ITAS...3163334L}%
  \BibitemOpen
  \bibfield  {author} {\bibinfo {author} {\bibfnamefont {Y.}~\bibnamefont
  {{Li}}}, \bibinfo {author} {\bibfnamefont {J.~E.}\ \bibnamefont
  {{Austermann}}}, \bibinfo {author} {\bibfnamefont {J.~A.}\ \bibnamefont
  {{Beall}}}, \bibinfo {author} {\bibfnamefont {S.~M.}\ \bibnamefont
  {{Bruno}}}, \bibinfo {author} {\bibfnamefont {S.~K.}\ \bibnamefont {{Choi}}},
  \bibinfo {author} {\bibfnamefont {N.~F.}\ \bibnamefont {{Cothard}}}, \bibinfo
  {author} {\bibfnamefont {K.~T.}\ \bibnamefont {{Crowley}}}, \bibinfo {author}
  {\bibfnamefont {S.~M.}\ \bibnamefont {{Duff}}}, \bibinfo {author}
  {\bibfnamefont {S.-P.~P.}\ \bibnamefont {{Ho}}}, \bibinfo {author}
  {\bibfnamefont {J.~E.}\ \bibnamefont {{Golec}}}, \bibinfo {author}
  {\bibfnamefont {G.~C.}\ \bibnamefont {{Hilton}}}, \bibinfo {author}
  {\bibfnamefont {M.}~\bibnamefont {{Hasselfield}}}, \bibinfo {author}
  {\bibfnamefont {J.}~\bibnamefont {{Hubmayr}}}, \bibinfo {author}
  {\bibfnamefont {B.~J.}\ \bibnamefont {{Koopman}}}, \bibinfo {author}
  {\bibfnamefont {M.}~\bibnamefont {{Lungu}}}, \bibinfo {author} {\bibfnamefont
  {J.}~\bibnamefont {{McMahon}}}, \bibinfo {author} {\bibfnamefont {M.~D.}\
  \bibnamefont {{Niemack}}}, \bibinfo {author} {\bibfnamefont {L.~A.}\
  \bibnamefont {{Page}}}, \bibinfo {author} {\bibfnamefont {M.}~\bibnamefont
  {{Salatino}}}, \bibinfo {author} {\bibfnamefont {S.~M.}\ \bibnamefont
  {{Simon}}}, \bibinfo {author} {\bibfnamefont {S.~T.}\ \bibnamefont
  {{Staggs}}}, \bibinfo {author} {\bibfnamefont {J.~R.}\ \bibnamefont
  {{Stevens}}}, \bibinfo {author} {\bibfnamefont {J.~N.}\ \bibnamefont
  {{Ullom}}}, \bibinfo {author} {\bibfnamefont {E.~M.}\ \bibnamefont
  {{Vavagiakis}}}, \bibinfo {author} {\bibfnamefont {Y.}~\bibnamefont
  {{Wang}}}, \bibinfo {author} {\bibfnamefont {E.~J.}\ \bibnamefont
  {{Wollack}}},\ and\ \bibinfo {author} {\bibfnamefont {Z.}~\bibnamefont
  {{Xu}}},\ }\href {https://doi.org/10.1109/TASC.2021.3063334} {\bibfield
  {journal} {\bibinfo  {journal} {IEEE Transactions on Applied
  Superconductivity}\ }\textbf {\bibinfo {volume} {31}},\ \bibinfo {eid}
  {3063334} (\bibinfo {year} {2021})},\ \Eprint
  {https://arxiv.org/abs/2101.02658} {arXiv:2101.02658 [astro-ph.IM]}
  \BibitemShut {NoStop}%
\end{thebibliography}%

\end{document}